\documentclass[english]{article}
\usepackage[T1]{fontenc}
\usepackage[latin9]{inputenc}
\usepackage{color}
\usepackage{babel}
\usepackage{mathrsfs}
\usepackage{amsmath}
\usepackage{amssymb}
\usepackage{graphicx}
\usepackage{setspace}
\usepackage{esint}
\usepackage[unicode=true,pdfusetitle,
 bookmarks=true,bookmarksnumbered=false,bookmarksopen=false,
 breaklinks=false,pdfborder={0 0 1},backref=false,colorlinks=false]
 {hyperref}

\makeatletter

\providecommand{\tabularnewline}{\\}

\pdfoutput=1

\usepackage{cite}

\usepackage{mathabx}

\makeatother

\begin{document}
\noindent \textbf{\Large{}From Local Chaos to Critical Slowing Down:
A Theory of the Functional Connectivity of Small Neural Circuits }\\
{\Large \par}

\noindent \begin{flushleft}
Diego Fasoli$^{1,\ast}$, Anna Cattani$^{1}$, Stefano Panzeri$^{1}$ 
\par\end{flushleft}

\medskip{}

\noindent \begin{flushleft}
\textbf{{1} Laboratory of Neural Computation, Center for Neuroscience
and Cognitive Systems @UniTn, Istituto Italiano di Tecnologia, 38068
Rovereto, Italy}
\par\end{flushleft}

\noindent \begin{flushleft}
\textbf{$\ast$ Corresponding Author. E-mail: diego.fasoli@iit.it}
\par\end{flushleft}

\section*{Abstract}

\noindent Functional connectivity is a fundamental property of neural
networks that quantifies the segregation and integration of information
between cortical areas. Due to mathematical complexity, a theory that
could explain how the parameters of mesoscopic networks composed of
a few tens of neurons affect the functional connectivity is still
to be formulated. Yet, many interesting problems in neuroscience involve
the study of networks composed of a small number of neurons. Based
on a recent study of the dynamics of small neural circuits, we combine
the analysis of local bifurcations of multi-population neural networks
of arbitrary size with the analytical calculation of the functional
connectivity. We study the functional connectivity in different regimes,
showing that external stimuli cause the network to switch from asynchronous
states characterized by weak correlation and low variability (local
chaos), to synchronous states characterized by strong correlations
and wide temporal fluctuations (critical slowing down). Local chaos
typically occurs in large networks, but here we show that it can also
be generated by strong stimuli in small neural circuits. On the other
side, critical slowing down is expected to occur when the stimulus
moves the network close to a local bifurcation. In particular, strongly
positive correlations occur at the saddle-node and Andronov-Hopf bifurcations
of the network, while strongly negative correlations occur when the
network undergoes a spontaneous symmetry-breaking at the branching-point
bifurcations. These results prove that the functional connectivity
of firing-rate network models is strongly affected by the external
stimuli even if the anatomical connections are fixed, and suggest
an effective mechanism through which biological networks can dynamically
modulate the encoding and integration of sensory information.

\section*{Author Summary}

\noindent Functional connectivity is nowadays one of the most debated
topics in neuroscience. It is a measure of the statistical dependencies
among neurons, from which we can infer the segregation and integration
of information in the nervous system. At the scale of cortical microcolumns,
the neural tissue is composed of circuits containing only a few tens
of neurons. However, and somewhat counter-intuitively, the functional
connectivity of small neural networks can be much more difficult to
study mathematically than that of large networks, because as is usual
with small numbers, at this scale of spatial organization statistical
procedures fail. For this reason, previous studies focused only on
the analysis of the functional connectivity in large-scale descriptions
of the neural tissue. In this work we introduce a new method for the
analysis of the functional connectivity of multi-population neural
networks of arbitrary size. In particular, we systematically quantify
how the functional connectivity is affected by the external stimuli.
Strong inputs drive the network toward asynchronous states where the
neurons are functionally disconnected. On the other side, for special
sets of stimuli the network becomes increasingly sensitive to external
perturbations. There the neural activity becomes synchronous and the
network is therefore characterized by strong functional integration.
This result suggests a possible neurophysiologic mechanism through
which sensory stimuli can dynamically modulate the information processing
capability of afferent cortical networks.

\section{Introduction \label{sec:Introduction}}

\noindent The brain is a complex organ with different scales of spatial
organization \cite{Sporns2006}. At the macroscopic scale, its complexity
is reflected by the high number of segregated sensory areas that accomplish
specialized information processing tasks, such as vision and audition.
Then, the sensory areas project to multimodal associative areas, where
information is integrated to provide us a coherent representation
of the world as we perceive it. This interplay of segregation and
integration has been proposed as a possible explanation of human complex
behavior \cite{Tononi1994} and is generally described by the term
\textit{functional connectivity} \cite{Rogers2007}. Therefore functional
connectivity underlies an information flow between and within cortical
areas, which at the macroscopic scale is measured by techniques such
as fMRI \cite{VanDenHeuvel2010}. More generally, Friston defines
the functional connectivity as the set of statistical dependencies
among neurons or neural populations \cite{Friston2011}. According
to this definition, the functional connectivity can be evaluated by
means of different theoretical measures \cite{David2004}, one of
the most used being cross-correlation. This allows us to define the
functional connectivity also at the mesoscopic scale, and to record
it by means of techniques such as EEG, MEG, ECoG and LFP \cite{He2008}.
The importance of the mesoscopic scale is underlined by its role in
shaping human cognitive functions. For example, LFP oscillations are
related to perceptual grouping and determine an exchange of information
between neighboring and distant cortical columns through neural synchronization
\cite{Singer1993}. Another example is represented by the columnar
mechanisms at the base of attention, which enhance the processing
of the relevant information of a complex environment and suppress
the unimportant one \cite{Harris2011}, resulting in a modulation
of the functional connectivity of the brain \cite{Cohen2009}.

Developing effective models of functional connectivity is nowadays
a central problem in theoretical neuroscience. Recently, the quantification
of the functional connectivity in terms of the cross-correlation structure
has been considered under different theoretical frameworks \cite{Bressloff2009,Renart2010,Pernice2011,Trousdale2012,Buice2013}.
An important aspect of the functional connectivity is to understand
how it is modulated by the most relevant parameters of the system,
in particular the stimulus and the strength of the synaptic connections,
since they are the most likely to change over time. In their pioneer
work \cite{Ginzburg1994}, Ginzburg and Sompolinsky developed a theory
of correlations for large neural networks described by Wilson-Cowan
equations. They proved that neural activity can switch from asynchronous
states characterized by weak correlation and low variability to synchronous
states characterized by strong correlations and wide temporal fluctuations.

Neurons are said to be in an \textit{asynchronous regime} when they
show uncorrelated activity \cite{Ecker2010,Renart2010,Tetzlaff2012},
while in mathematics a regime characterized by independent (though
interacting) units is called \textit{local chaos} \footnote{Local chaos is known in the kinetic theory of gases as \textit{molecular
chaos}. It was originally introduced by Boltzmann with the name \textit{stosszahlansatz}
(collision-number hypothesis) in his studies on the second law of
thermodynamics \cite{Boltzmann1872}. According to this hypothesis,
particles in a gas are statistically independent, even if they interact
with each other. However, intuitively after a collision the particles
should not be independent anymore since they exchange information.
Indeed it can be proven that the inter-particle dependence never vanish
during time evolution in a system composed of a finite number of particles
and that the Boltzmann's hypothesis is true only in the ideal limit
of infinitely many particles (the so called \textit{thermodynamic
limit}).}. For the sake of clarity, it is important to observe that asynchrony
is a weaker condition than local chaos. Indeed, for a general probability
distribution, independence does not imply decorrelation. However,
the two conditions are equivalent if the neurons are jointly normally
distributed, as in the case of the theory we propose in this work.
In \cite{Ginzburg1994} the authors proved that asynchronous states
occur in large networks since the correlations between neurons vanish
as $\frac{1}{N}$, where $N$ is the size of the network. In a similar
way, the emergence of local chaos in large neural networks was proven
in \cite{Samuelides2007,Touboul2012,Baladron2012a,Baladron2012b}.
Indeed, independence between interacting units is usually the hypothesis
invoked to justify the mean-field approximation of large systems.
On the other side, a \textit{synchronous regime} typically occurs
when the network undergoes critical slowing down. Generally this phenomenon
happens when a system becomes increasingly sensitive to external perturbations
\cite{Kefi2013}. In this situation the state variables undergo large
and asymmetric fluctuations, with a strong increase of the cross-
and auto-correlation functions \cite{Scheffer2009,Kuehn2013}. Critical
slowing down usually occurs at the bifurcation points of the system
(but not all of them), where small variations of the parameters cause
qualitative changes in its dynamics. For example, in \cite{Ginzburg1994}
the authors showed the formation of critical slowing down in large
networks when they approach a saddle-node or a Andronov-Hopf bifurcation,
namely before catastrophic transitions or the emergence of oscillatory
activity respectively.

Current theories of correlation can be typically applied to networks
composed of few thousands of neurons or more, which represent the
upper limit of the mesoscopic scale. However, and somewhat counter-intuitively,
the cross-correlation structure of small neural networks containing
only a few tens of neurons can be much more difficult to study mathematically
than that of large networks. This is mainly due to the impossibility
to apply the powerful methods of statistical analysis, such as the
\textit{law of large numbers} and the \textit{central limit theorem},
to small neural circuits. Indeed, these statistical techniques can
be typically applied in the limit of large populations of independent
neurons. However, in \cite{Fasoli2016} we recently introduced a method
for studying the dynamics of neural circuits of arbitrary size, which
does not rely on statistical methods. This approach proved effective
in describing analytically the local bifurcations of small networks
composed of a few tens of neurons such as cortical microcolumns \cite{Mountcastle1997},
which represent the lower bound of the mesoscopic scale.

In this work we stochastically perturb a generalized version of the
deterministic firing-rate network model introduced in \cite{Fasoli2016}.
This allows us to develop a theory of the functional connectivity
of small neural circuits composed of several populations. Similarly
to \cite{Ginzburg1994}, we find that such networks display both synchronous
and asynchronous regimes, with important qualitative and quantitative
differences. As in \cite{Fasoli2015}, we prove that local chaos may
occur also in small networks for strongly depolarizing or strongly
hyperpolarizing stimuli. Then, as in \cite{Ginzburg1994}, we prove
the emergence of critical slowing down at the saddle-node and Andronov-Hopf
bifurcations of the network, but we extend this result to the case
of neural circuits of arbitrary size and with arbitrary correlations
between the stochastic perturbations. Moreover, in \cite{Fasoli2016}
we found that small networks undergo also special bifurcations known
as \textit{branching points} or \textsl{pitchfork bifurcations}. Branching
points correspond to a \textit{spontaneous symmetry-breaking} of the
neural activity. There we observe the spontaneous formation of heterogeneous
activity from homogeneous inhibitory neurons without explicit asymmetries
in the neural equations. In this work we prove that at these special
bifurcation points the activity between inhibitory neurons undergoes
critical slowing down characterized by strong \textit{anti-correlation}:
this is a consequence of the broken symmetry of the network, that
was not considered in \cite{Ginzburg1994}. 

A systematic analysis of critical slowing down at bifurcations up
to codimension two can be found in \cite{Kuehn2013}. Kuehn's analysis
is based on the \textit{normal forms} of bifurcations, namely simple
dynamical systems which are locally equivalent to all systems exhibiting
those bifurcations \cite{Kuznetsov1998}. For this reason, the mathematical
analysis in \cite{Kuehn2013} is very general. However, the essential
consequences of the theory have not been explicitly formulated in
the specific case of neural networks. Therefore quantifying the relation
between critical slowing down and the parameters of small neural networks
is still an open problem, that we tackle in this work. Our approach
is based on linear algebra, not on normal forms, thus it is accessible
also to less mathematically minded readers.

The article is organized as follows. In Sec.~(\ref{sec:Materials-and-Methods})
we introduce the firing-rate network model that we use for the calculation
of the functional connectivity (SubSec.~(\ref{sub:The-Firing-Rate-Network-Model})),
and different measures of functional connectivity in terms of the
cross-correlation between neurons or neural populations (SubSec.~(\ref{sub:Measures-of-Functional-Connectivity})).
In Sec.~(\ref{sec:Results}) we compare analytical and numerical
calculations of the cross-correlation in the special case of networks
composed of two populations. In more detail, in SubSec.~(\ref{sub:Local-Chaos})
we prove the formation of local chaos for strong stimuli, while in
SubSec.~(\ref{sub:Critical-Slowing-Down}) we show the emergence
of critical slowing down at the saddle-node, Andronov-Hopf and branching-point
bifurcations of the network. To conclude, in Sec.~(\ref{sec:Discussion})
we discuss the importance and the biological implications of our results,
while the extension of the theory to the case of an arbitrary number
of neural populations has been developed in the Supplementary Materials.

\section{Materials and Methods \label{sec:Materials-and-Methods}}

In this section we introduce the multi-population network that we
study in the article (SubSec.~(\ref{sub:The-Firing-Rate-Network-Model})).
Moreover, we propose different measures of functional connectivity
that can be used to compare the theory with the numerical simulations
(SubSec.~(\ref{sub:Measures-of-Functional-Connectivity})).

\subsection{The Firing-Rate Network Model \label{sub:The-Firing-Rate-Network-Model}}

Similarly to \cite{Fasoli2016}, we make some assumptions in order
to make the network analytically tractable. In particular, we assume
that the neurons in each population have homogeneous parameters, that
the neurons are all-to-all connected to each other, and that the axonal
delays are negligible. Moreover, we describe random fluctuations in
the network by means of a white noise component in the external stimuli.
Indeed, in humans and many other vertebrates, white noise may be interpreted
as an ongoing flux of non-specific excitatory input from the reticular
activating system within the brainstem \cite{Steyn-Ross2004}.

In more detail, we describe the network by means of the following
system of stochastic differential equations:

\begin{spacing}{0.8}
\begin{center}
{\small{}
\begin{equation}
\frac{dV_{i}\left(t\right)}{dt}=-\frac{1}{\tau_{i}}V_{i}\left(t\right)+\frac{1}{M_{i}}\sum_{j=0}^{N-1}J_{ij}\mathscr{A}_{j}\left(V_{j}\left(t\right)\right)+I_{i}\left(t\right)+\sigma_{i}^{\mathscr{B}}\frac{d\mathscr{B}_{i}\left(t\right)}{dt},\quad i=0,...,N-1.\label{eq:exact-rate-equations}
\end{equation}
}
\par\end{center}{\small \par}
\end{spacing}

\noindent \begin{flushleft}
Eq.~(\ref{eq:exact-rate-equations}) represents the stochastic perturbation
to the firing-rate network model discussed in \cite{Fasoli2016}.
$N$\textcolor{blue}{{} }is the number of neurons in the network, $V_{i}\left(t\right)$
is the membrane potential of the $i$th neuron at the time instant
$t$, and $\tau_{i}$ is its membrane time constant. The normalization
factor $M_{i}$ represents the number of incoming connections to the
$i$th neuron, while $J_{ij}$ is the weight of the synaptic connection
from the $j$th (presynaptic) neuron to the $i$th (postsynaptic)
neuron. $\mathscr{A}_{j}\left(\cdot\right)$ is an algebraic activation
function which converts the membrane potential $V$ into the corresponding
firing rate $\nu=\mathscr{A}\left(V\right)$ according to the formula:
\par\end{flushleft}

\begin{spacing}{0.8}
\begin{center}
{\small{}
\begin{equation}
\mathscr{A}_{j}\left(V\right)=\frac{\nu_{j}^{\mathrm{max}}}{2}\left[1+\frac{\frac{\Lambda_{j}}{2}\left(V-V_{j}^{T}\right)}{\sqrt{1+\frac{\Lambda_{j}^{2}}{4}\left(V-V_{j}^{T}\right)^{2}}}\right].\label{eq:algebraic-activation-function}
\end{equation}
}
\par\end{center}{\small \par}
\end{spacing}

\noindent Here $\nu_{j}^{\mathrm{max}}$ is the maximum firing rate
of the neuron, $V_{j}^{T}$ is the threshold of the activation function,
and $\Lambda_{j}$ is its slope parameter. The latter represents the
``speed'' with which the neuron switches between low rates ($\nu_{j}\approx0$)
and high rates ($\nu_{j}\approx\nu_{j}^{\mathrm{max}}$). Moreover,
in Eq.~(\ref{eq:exact-rate-equations}) $I_{i}\left(t\right)$ is
a deterministic external input (i.e. the stimulus) to the $i$th neuron,
while $\sigma_{i}^{\mathscr{B}}\frac{d\mathscr{B}_{i}\left(t\right)}{dt}$
is a white noise input with normal distribution and standard deviation
$\sigma_{i}^{\mathscr{B}}\ll1$. The functions $\mathscr{B}_{i}\left(t\right)$
are arbitrarily correlated Brownian motions, which represent the source
of stochasticity of the model.

As in \cite{Fasoli2016}, in order to make our analysis analytically
tractable, we suppose that all the parameters of the system are indexed
only at the population level. This means that within a given population
the parameters are homogeneous (see \cite{Fasoli2016} for a discussion
about the effects of heterogeneity). In other terms, this hypothesis
allows us to define a modular network which is composed of an arbitrary
number $\mathfrak{P}$ of homogeneous neural communities or populations.
We define $N_{\alpha}$ to be the size of population $\alpha$ (namely
the number of neurons within that population), with ${\displaystyle \sum_{\alpha=0}^{\mathfrak{P}-1}N_{\alpha}}=N$,
and we rearrange the neurons so that the structural connectivity of
the network can be written as follows:

\begin{spacing}{0.8}
\begin{center}
{\small{}
\begin{equation}
\begin{array}{ccc}
J=\left[\begin{array}{cccc}
\mathfrak{J}_{00} & \mathfrak{J}_{01} & \cdots & \mathfrak{J}_{0,\mathfrak{P}-1}\\
\mathfrak{J}_{10} & \mathfrak{J}_{11} & \cdots & \mathfrak{J}_{1,\mathfrak{P}-1}\\
\vdots & \vdots & \ddots & \vdots\\
\mathfrak{J}_{\mathfrak{P}-1,0} & \mathfrak{J}_{\mathfrak{P}-1,1} & \cdots & \mathfrak{J}_{\mathfrak{P}-1,\mathfrak{P}-1}
\end{array}\right], &  & \mathfrak{J}_{\alpha\beta}=\begin{cases}
J_{\alpha\alpha}\left(\mathbb{I}_{N_{\alpha}}-\mathrm{Id}_{N_{\alpha}}\right), & \;\mathrm{for}\;\alpha=\beta\\
\\
J_{\alpha\beta}\mathbb{I}_{N_{\alpha},N_{\beta}}, & \;\mathrm{for}\;\alpha\neq\beta
\end{cases}\end{array}\label{eq:synaptic-connectivity-matrix}
\end{equation}
}
\par\end{center}{\small \par}
\end{spacing}

\noindent \begin{flushleft}
for $\alpha,\beta=0,\ldots,\mathfrak{P}-1$. The real numbers $J_{\alpha\beta}$
are free parameters that describe the strength of the synaptic connections
from the population $\beta$ to the population $\alpha$. We have
$J_{\alpha\beta}\geq0$ $\forall\alpha$ if the population $\beta$
is excitatory, and $J_{\alpha\beta}\leq0$ $\forall\alpha$ if it
is inhibitory. Moreover, $\mathbb{I}_{N_{\alpha},N_{\beta}}$ is the
$N_{\alpha}\times N_{\beta}$ all-ones matrix (here we use the simplified
notation $\mathbb{I}_{N_{\alpha}}\overset{\mathrm{def}}{=}\mathbb{I}_{N_{\alpha},N_{\alpha}}$),
while $\mathrm{Id}_{N_{\alpha}}$ is the $N_{\alpha}\times N_{\alpha}$
identity matrix. From our assumption on the indexes, we also obtain
that the external input currents are organized into $\mathfrak{P}$
vectors $\boldsymbol{I}_{\alpha}$, one for each population, and such
that:
\par\end{flushleft}

\begin{spacing}{0.8}
\begin{center}
{\small{}
\[
\boldsymbol{I}_{\alpha}\left(t\right)=I_{\alpha}\left(t\right)\boldsymbol{1}_{N_{\alpha}},
\]
}
\par\end{center}{\small \par}
\end{spacing}

\noindent where $\boldsymbol{1}_{N_{\alpha}}\overset{\mathrm{def}}{=}\mathbb{I}_{N_{\alpha},1}$
is the $N_{\alpha}\times1$ all-ones vector. The same subdivision
between populations is performed for the parameters $M$, $\tau$,
$\nu^{\mathrm{max}}$, $\Lambda$, $V_{T}$.

We also suppose that the correlation structure of the white noise
$\frac{d\mathscr{B}_{i}\left(t\right)}{dt}$ is given by the matrix
$\Sigma^{\mathscr{B}}=\left[\varSigma_{\alpha\beta}^{\mathscr{B}}\right]_{\forall\left(\alpha,\beta\right)}$,
where:

\begin{spacing}{0.8}
\begin{center}
{\small{}
\begin{equation}
\varSigma_{\alpha\beta}^{\mathscr{B}}=\begin{cases}
\left(\sigma_{\alpha}^{\mathscr{B}}\right)^{2}\left[\mathrm{Id}_{N_{\alpha}}+C_{\alpha\alpha}^{\mathscr{B}}\left(\mathbb{I}_{N_{\alpha}}-\mathrm{Id}_{N_{\alpha}}\right)\right], & \;\mathrm{for}\;\alpha=\beta\\
\\
\sigma_{\alpha}^{\mathscr{B}}\sigma_{\beta}^{\mathscr{B}}C_{\alpha\beta}^{\mathscr{B}}\mathbb{I}_{N_{\alpha},N_{\beta}}, & \;\mathrm{for}\;\alpha\neq\beta
\end{cases}\label{eq:noise-covariance-matrix}
\end{equation}
}
\par\end{center}{\small \par}
\end{spacing}

\noindent and $C_{\alpha\beta}^{\mathscr{B}}=C_{\beta\alpha}^{\mathscr{B}}$
since $\Sigma^{\mathscr{B}}$ must be symmetric in order to be a true
covariance matrix. $\Sigma^{\mathscr{B}}$ determines the correlation
structure of the white noise since $\mathrm{Cov}\left(\frac{d\mathscr{B}_{i}\left(t\right)}{dt},\frac{d\mathscr{B}_{j}\left(s\right)}{ds}\right)=\left[\Sigma^{\mathscr{B}}\right]_{ij}\delta\left(t-s\right)$.

\begin{figure}
\begin{centering}
\includegraphics[scale=0.33]{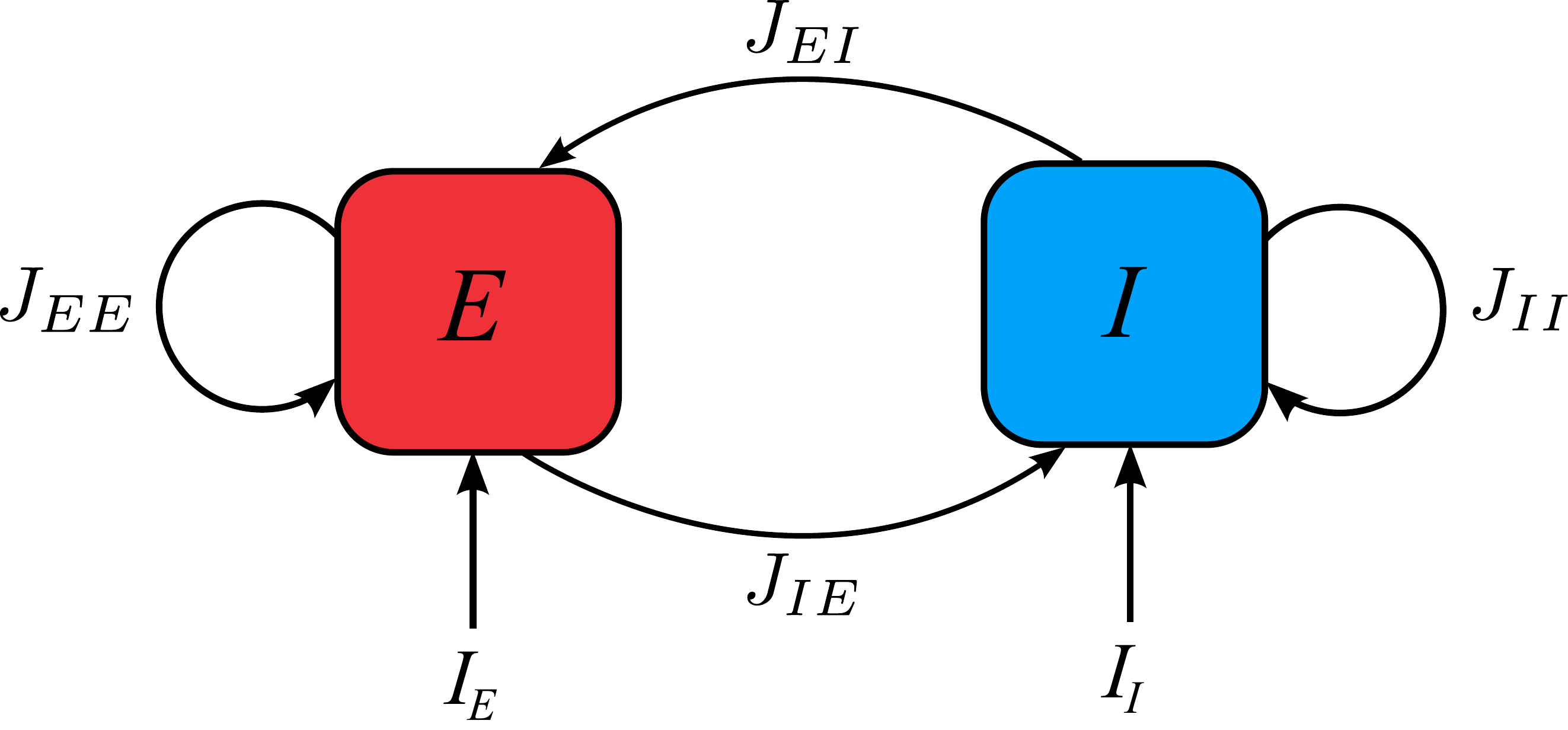}
\par\end{centering}

\protect\caption{\label{Fig:network-structure} \small \textbf{Example of neural network
for $\boldsymbol{\mathfrak{P}=2}$.} The two populations, one excitatory
($E$) and one inhibitory ($I$), are composed of fully-connected
neurons.}
\end{figure}

\begin{figure}
\begin{centering}
\includegraphics[scale=0.7]{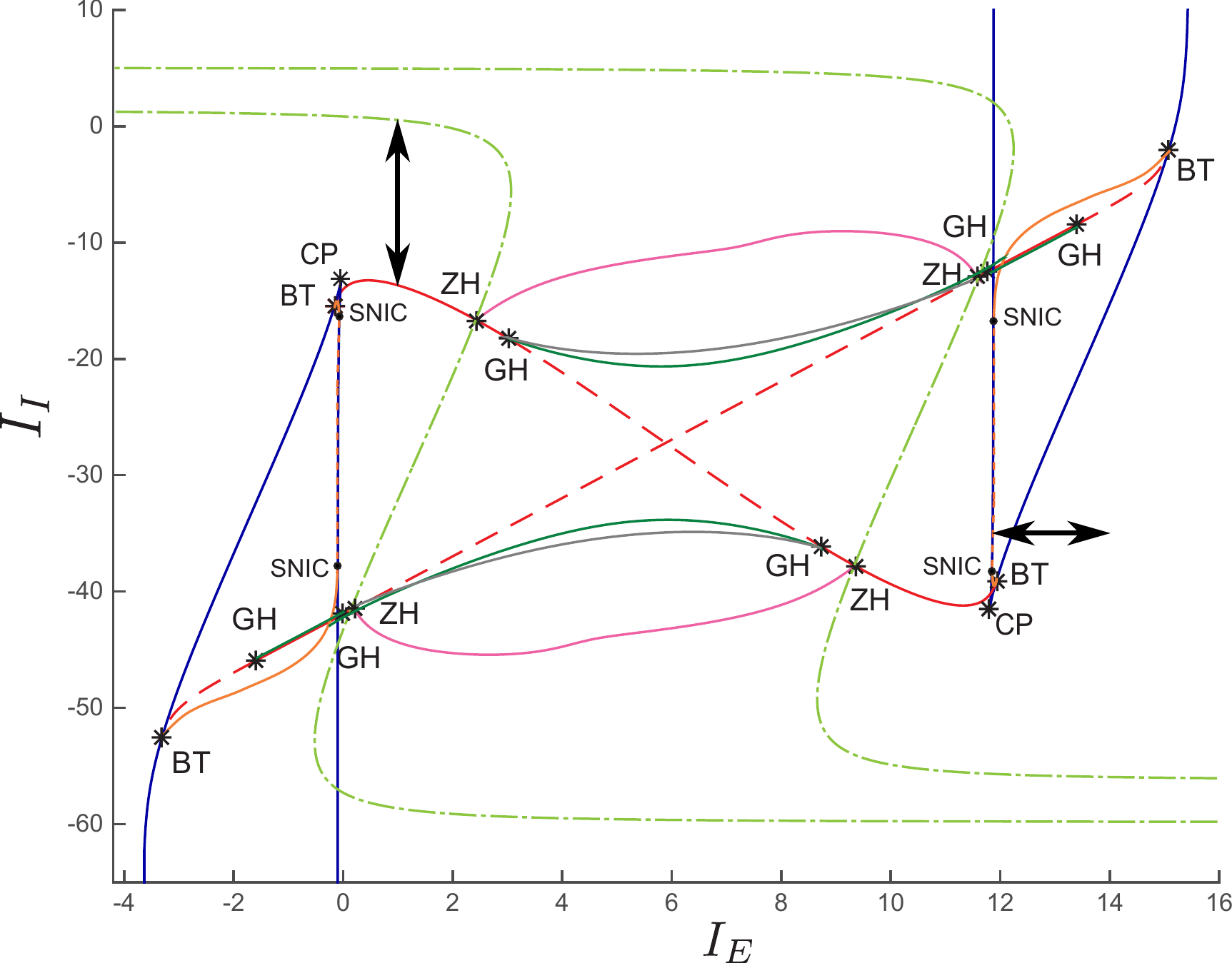}
\par\end{centering}

\protect\caption{\label{Fig:codimension-two-bifurcation-diagram} \small \textbf{Codimension
two bifurcation diagram in the $\boldsymbol{I_{E}-I_{I}}$ plane for
$\boldsymbol{\mathfrak{P}=2}$.} This diagram was obtained in \cite{Fasoli2016}
for the values of the parameters reported in Tab.~(\ref{tab:network-parameters}).
The blue curves represent the saddle-node bifurcations (LP for short
in Figs. (\ref{fig:local-chaos-and-saddle-node-bifurcation}), (\ref{fig:functional-connectivity-for-correlated-Brownian-motions}))
on the primary branch of stationary solutions of Eq.~(\ref{eq:exact-rate-equations}),
with cusp bifurcations (CP). The red curves correspond to the Andronov-Hopf
bifurcations (H for short in Figs. (\ref{fig:Andronov-Hopf-and-branching-point-bifurcations}),
(\ref{fig:functional-connectivity-for-correlated-Brownian-motions}))
on the primary branch, which in turn are divided into supercritical
(plain) and subcritical (dashed) portions. The supercritical/subcritical
portions are bounded by a generalized Hopf bifurcation (GH), and Bogdanov-Takens
bifurcations (BT). The latter are the contact points among saddle-node,
Andronov-Hopf and homoclinic bifurcation curves on the primary branch
(hyperbolic-saddle/saddle-node homoclinic bifurcations are represented
by plain/dashed orange curves). Saddle-node on invariant circle bifurcations
(SNIC) correspond to the contact points between the saddle-node and
the homoclinic curves. GH generates limit point of cycles curves,
represented by dark green lines, that collapse into the homoclinic
curves. The gray lines represent the torus bifurcations, while the
light green dot-dashed curves correspond to the branching-point bifurcations
(BP for short in Figs. (\ref{fig:local-chaos-and-saddle-node-bifurcation}),
(\ref{fig:Andronov-Hopf-and-branching-point-bifurcations}), (\ref{fig:functional-connectivity-for-correlated-Brownian-motions})).
The purple curves represent the Andronov-Hopf bifurcations that originate
from the secondary branches, which meet the branching-point curves
and the other Andronov-Hopf curves at the zero-Hopf bifurcations (ZH).
The double-headed black arrows represent the ranges in which we varied
the stimuli $I_{E,I}$ in order to study the behavior of the functional
connectivity. In more detail, on the horizontal arrow the network
switches from local chaos to critical slowing down near a saddle-node
bifurcation (see also Fig.~(\ref{fig:local-chaos-and-saddle-node-bifurcation})).
Moreover, on the vertical arrow the network switches from positively
correlated activity at the Andronov-Hopf bifurcation curve, to anti-correlated
activity in the inhibitory population at the branching-point curve
(see also Fig.~(\ref{fig:Andronov-Hopf-and-branching-point-bifurcations})).
Adapted from \cite{Fasoli2016} with permission of the authors.}
\end{figure}

A cortical column can be thought of as a network of neural masses
distributed vertically across layers, and therefore it is composed
of several populations of excitatory and inhibitory neurons (see for
example \cite{Binzegger2004}). Our theory can be used to study such
cortical architectures, but the complexity of the resulting formulas
increases considerably with the number of populations. Thus for the
sake of example, we focus on the case $\mathfrak{P}=2$ with one excitatory
($E$) and one inhibitory ($I$) neural population, which is commonly
considered a good approximation of a single neural mass \cite{Grimbert2008}.
The case of networks with an arbitrary number of populations is considered
in the Supplementary Materials. From now on, it is convenient to change
slightly the notation, and to consider $\alpha,\beta=E,I$ rather
than $\alpha,\beta=0,1$ (see Fig.~(\ref{Fig:network-structure})).
 Since we study the case of two neural populations, we can take advantage
of the detailed bifurcation analysis performed in \cite{Fasoli2016}
(see also Fig.~(\ref{Fig:codimension-two-bifurcation-diagram})),
which we will use to determine where the functional connectivity undergoes
the most interesting variations. 
 
\begin{table}
\begin{centering}
{\small{}}%
\begin{tabular}{|c|c|c|c|c|}
\hline 
\textbf{\small{}Population Sizes} & \textbf{\small{}Synaptic Weights} & \textbf{\small{}Activation Functions} & \textbf{\small{}Brownian Motions} & \textbf{\small{}Other}\tabularnewline
\hline 
{\small{}$N_{E}=8$} & {\small{}$J_{EE}=10$} & {\small{}$\nu_{E}^{\mathrm{max}}=\nu_{I}^{\mathrm{max}}=1$} & {\small{}$\sigma_{E}^{\mathscr{B}}=\sigma_{I}^{\mathscr{B}}=10^{-4}$} & {\small{}$\tau_{E}=\tau_{I}=1$}\tabularnewline
{\small{}$N_{I}=2$} & {\small{}$J_{EI}=-70$} & {\small{}$\Lambda_{E}=\Lambda_{I}=2$} & {\small{}$C_{EE}^{\mathscr{B}}=C_{II}^{\mathscr{B}}=C_{EI}^{\mathscr{B}}=0$} & \tabularnewline
 & {\small{}$J_{IE}=70$} & {\small{}$V_{E}^{T}=V_{I}^{T}=2$} &  & \tabularnewline
 & {\small{}$J_{II}=-34$} &  &  & \tabularnewline
\hline 
\end{tabular}
\par\end{centering}{\small \par}

\protect\caption{\label{tab:network-parameters} \small \textbf{Values of the parameters
of the network for $\boldsymbol{\mathfrak{P}=2}$.} These parameters
have been used to generate all the figures in the article, but Fig.~(\ref{fig:functional-connectivity-for-correlated-Brownian-motions}),
where the functional connectivity is evaluated for different values
of $C_{\alpha\beta}^{\mathscr{B}}$. The ratio $\frac{N_{E}}{N_{I}}=4$
reflects the proportion between excitatory and inhibitory neurons
in biological circuits (see \cite{Markram2004}). Our theory can be
applied to networks of arbitrary size, but in the article we consider
the case of small networks ($N=10$ in this example), see text.}
\end{table}

\subsection{Measures of Functional Connectivity \label{sub:Measures-of-Functional-Connectivity}}

Cross-correlation is one of the most studied measures of functional
connectivity \cite{David2004}. For simplicity here we focus on pairwise
correlations, which can be calculated through the Pearson coefficient
formula:

\begin{spacing}{0.80000000000000004}
\begin{center}
{\small{}
\begin{equation}
\mathrm{Corr}\left(V_{i}\left(t\right),V_{j}\left(t\right)\right)\overset{\mathrm{def}}{=}\frac{\mathrm{Cov}\left(V_{i}\left(t\right),V_{j}\left(t\right)\right)}{\sqrt{\mathrm{Var}\left(V_{i}\left(t\right)\right)\mathrm{Var}\left(V_{j}\left(t\right)\right)}},\label{eq:pairwise-correlation}
\end{equation}
}
\par\end{center}{\small \par}
\end{spacing}

\noindent where $\mathrm{Var}\left(V_{i}\left(t\right)\right)=\mathrm{Cov}\left(V_{i}\left(t\right),V_{i}\left(t\right)\right)$.
In \cite{Fasoli2015} the authors derived the analytical expression
of the covariance of the rate model (\ref{eq:exact-rate-equations})
for a generic connectivity matrix $J$. This formula reads:

\begin{spacing}{0.8}
\noindent \begin{center}
{\small{}
\begin{equation}
\mathrm{Cov}\left(V_{i}\left(t\right),V_{j}\left(t\right)\right)=\sum_{k=0}^{N-1}\left(\sigma_{k}^{\mathscr{B}}\right)^{2}\int_{0}^{t}\Phi_{ik}\left(t-s\right)\Phi_{jk}\left(t-s\right)ds,\label{eq:pairwise-covariance}
\end{equation}
}
\par\end{center}{\small \par}
\end{spacing}

\noindent where $\Phi\left(t\right)=e^{\mathcal{J}t}$ is the fundamental
matrix of the system at time $t$, while $\mathcal{J}$ is its Jacobian
matrix (which depends on $J$). However, when applied to our connectivity
matrix (see Eq.(\ref{eq:synaptic-connectivity-matrix}) for $\mathfrak{P}=2$),
Eqs.~(\ref{eq:pairwise-correlation}) + (\ref{eq:pairwise-covariance})
provide a very cumbersome expression of the cross-correlation. Thus,
for simplicity, in this article we consider only the limit $t\rightarrow+\infty$,
even if correlations may be calculated at any finite $t$, if desired.
Moreover, in Sec.~(\ref{sec:Results}) we will compare the resulting
analytical expression with numerical evaluations of the correlation.
In particular, the numerical results are obtained by integrating the
neural equations (\ref{eq:exact-rate-equations}) with the Euler-Maruyama
method, for the values of the parameters reported in Tab.~(\ref{tab:network-parameters}).
The integration time step is $\Delta t=0.001$, and the equations
are integrated with a Monte Carlo method over $5,000$ repetitions
of the network dynamics in the temporal interval $t=\left[0,30\right]$.
We assume that at $t=30$ the transient regime of the correlation
has already passed (so that correlation has already converged to its
equilibrium solution), which is confirmed by the good agreement between
the analytical and numerical results. According to \cite{Fasoli2015},
in order to compare analytical and numerical approximations of the
functional connectivity, the membrane potentials have to stay as close
as possible to a given equilibrium point. In order to avoid jumps
of the potentials between different equilibria when the network is
close to a saddle-node bifurcation, we consider Brownian motions with
small standard deviation, namely $\sigma_{E}^{\mathscr{B}}=\sigma_{I}^{\mathscr{B}}=10^{-4}$.
Moreover, this choice alleviates another numerical issue, as described
hereafter. When the network is close to an Andronov-Hopf bifurcation,
two eigenvalues are complex conjugate, therefore they give rise to
a focus with damped oscillations in the phase portrait. This means
that the random fluctuations of the noise move the state of the network
from its equilibrium point, causing undesired sustained oscillations
whose frequency corresponds to the imaginary part of the eigenvalues
of the Jacobian matrix \cite{Wallace2011}. Only small $\sigma_{E,I}^{\mathscr{B}}$
prevent the formation of wide oscillations around the equilibrium
solution. On the other side, when the network is far from a bifurcation
point, we obtain a good agreement between analytical and numerical
results also for Brownian motions with larger standard deviations,
namely $\sigma_{E,I}^{\mathscr{B}}\sim10^{-1}$ (results not shown).
For even larger standard deviations, higher-order corrections to our
perturbative approach must be considered, but this is beyond the purpose
of the article.

Cross-correlation is related to the underlying information flow between
neurons by the formula of the mutual information:

\begin{spacing}{0.8}
\begin{center}
{\small{}
\begin{equation}
\mathcal{I}_{ij}\left(t\right)\overset{\mathrm{def}}{=}\int_{\mathbb{R}^{2}}p\left(V_{i},V_{j},t\right)\log\left(\frac{p\left(V_{i},V_{j},t\right)}{p\left(V_{i},t\right)p\left(V_{j},t\right)}\right)dV_{i}dV_{j}=-\frac{1}{2}\log\left(1-\mathrm{Corr}^{2}\left(V_{i}\left(t\right),V_{j}\left(t\right)\right)\right),\label{eq:mutual-information}
\end{equation}
}
\par\end{center}{\small \par}
\end{spacing}

\noindent where $p\left(V_{i},V_{j},t\right)$ is the $2$-neurons
joint probability density at time $t$, while $p\left(V_{i},t\right)=\int_{-\infty}^{+\infty}p\left(V_{i},V_{j},t\right)dV_{j}$
is the corresponding $1$-neuron density. We observe that the last
identity in Eq.~(\ref{eq:mutual-information}) holds only for normal
probability distributions. This is indeed our case, since we are going
to adopt a linear approximation of Eq.~(\ref{eq:exact-rate-equations}),
which is justified by our assumption $\sigma^{\mathscr{B}}\ll1$.
Eq.~(\ref{eq:mutual-information}) shows that the mutual information
$\mathcal{I}_{ij}\left(t\right)$ depends trivially on the pairwise
correlation between neurons. In particular, $\mathrm{Corr}\left(V_{i}\left(t\right),V_{j}\left(t\right)\right)\rightarrow0$
implies $\mathcal{I}_{ij}\left(t\right)\rightarrow0$ (functional
disconnection), while $\mathrm{Corr}\left(V_{i}\left(t\right),V_{j}\left(t\right)\right)\rightarrow\pm1$
implies $\mathcal{I}_{ij}\left(t\right)\rightarrow\infty$ (functional
integration).

However, neuroscientists make use of measures of correlation between
firing rates, rather than between membrane potentials. This is because
only spiking events are transmitted to other neurons, while subthreshold
membrane fluctuations are not. Since in our model the firing rates
are given by the relation $\nu=\mathscr{A}\left(V\right)$, we get
:

\begin{spacing}{0.8}
\begin{center}
{\small{}
\begin{equation}
\mathrm{Corr}\left(\nu_{i}\left(t\right),\nu_{j}\left(t\right)\right)\approx\mathrm{Corr}\left(V_{i}\left(t\right),V_{j}\left(t\right)\right),\label{eq:correlation-between-firing-rates}
\end{equation}
}
\par\end{center}{\small \par}
\end{spacing}

\noindent under the assumption $\sigma^{\mathscr{B}}\ll1$, as proven
in \cite{Fasoli2015}. Still, the network is described by voltage-based
equations (see (\ref{eq:exact-rate-equations})), thus it is more
natural to study the correlation structure between the membrane potentials
and to get that between the firing rates accordingly. 

Furthermore, it is also possible to calculate the cross-correlation
between mesoscopic quantities. For example, we introduce the \textit{neural
activity} of a group of neurons \cite{Gerstner2002}:

\begin{spacing}{0.8}
\begin{center}
{\small{}
\[
a_{\mathcal{G}}\left(t\right)\overset{\mathrm{def}}{=}\frac{1}{N_{\mathcal{G}}}\sum_{i\in\mathcal{G}}\nu_{i}\left(t\right),
\]
}
\par\end{center}{\small \par}
\end{spacing}

\noindent where the sum is over all the neurons in a given group $\mathcal{G}$
of size $N_{\mathcal{G}}$. The neural activity can be interpreted
intuitively by observing that, in the limit of large $\Lambda$ (see
Eq.~(\ref{eq:algebraic-activation-function})), $a_{\mathcal{G}}\left(t\right)$
corresponds to the fraction of firing neurons in $\mathcal{G}$. Indeed,
$\underset{\Lambda\rightarrow\infty}{\lim}\mathscr{A}\left(V\right)=\nu^{\mathrm{max}}H\left(V-V^{T}\right)$,
where $H\left(\cdot\right)$ is the Heaviside step function. Therefore
$\underset{\Lambda\rightarrow\infty}{\lim}a_{\mathcal{G}}\left(t\right)=\frac{\nu^{\mathrm{max}}}{N_{\mathcal{G}}}N_{\mathrm{fire}}\left(t\right)$,
where $N_{\mathrm{fire}}\left(t\right)=\sum_{i\in\mathcal{G}}H\left(V-V^{T}\right)$
is the number of firing neurons (i.e. such that $V>V^{T}$) in $\mathcal{G}$
at the time instant $t$. The correlation between the activities of
two neural groups $\mathcal{G}$ and $\mathcal{H}$ is:

\begin{spacing}{0.8}
\begin{center}
{\small{}
\begin{align}
\mathrm{Corr}\left(a_{\mathcal{G}}\left(t\right),a_{\mathcal{H}}\left(t\right)\right) & =\frac{\sum_{i\in\mathcal{G},j\in\mathcal{H}}\mathrm{Cov}\left(\nu_{i}\left(t\right),\nu_{j}\left(t\right)\right)}{\sqrt{\left[\sum_{i,j\in\mathcal{G}}\mathrm{Cov}\left(\nu_{i}\left(t\right),\nu_{j}\left(t\right)\right)\right]\left[\sum_{i,j\in\mathcal{H}}\mathrm{Cov}\left(\nu_{i}\left(t\right),\nu_{j}\left(t\right)\right)\right]}}\nonumber \\
\label{eq:correlation-between-neural-activities}\\
 & =\frac{N_{\mathcal{G}}N_{\mathcal{H}}\mathrm{Corr}\left(\mathcal{G},H\right)}{\sqrt{\left[N_{\mathcal{G}}+\left(N_{\mathcal{G}}^{2}-N_{\mathcal{G}}\right)\mathrm{Corr}\left(\mathcal{G},\mathcal{G}\right)\right]\left[N_{\mathcal{H}}+\left(N_{\mathcal{H}}^{2}-N_{\mathcal{H}}\right)\mathrm{Corr}\left(\mathcal{H},\mathcal{H}\right)\right]}},\nonumber 
\end{align}
}
\par\end{center}{\small \par}
\end{spacing}

\noindent where we defined $\mathrm{Corr}\left(\mathcal{G},\mathcal{H}\right)\overset{\mathrm{def}}{=}\left.\mathrm{Corr}\left(\nu_{i}\left(t\right),\nu_{j}\left(t\right)\right)\right|_{i\in\mathcal{G},j\in\mathcal{H}}$
and $\mathrm{Corr}\left(\mathcal{G},\mathcal{G}\right)\overset{\mathrm{def}}{=}\left.\mathrm{Corr}\left(\nu_{i}\left(t\right),\nu_{j}\left(t\right)\right)\right|_{i,j\in\mathcal{G},\,i\neq j}$
(similarly for $\mathcal{H}$). The last equality of Eq.~(\ref{eq:correlation-between-neural-activities})
holds only if $\mathcal{G}$ is a subset of a neural population where
spontaneous symmetry-breaking did not occur, and the same for $\mathcal{H}$
(so for example if the neurons in $\mathcal{G}$ are excitatory while
those in $\mathcal{H}$ are inhibitory, in the case $\mathfrak{P}=2$
with weak inhibition). Therefore from Eqs.~(\ref{eq:correlation-between-firing-rates})
+ (\ref{eq:correlation-between-neural-activities}) we observe that
also the correlation between population activities can be expressed
in terms of the correlation between the corresponding membrane potentials.

\section{Results \label{sec:Results}}

In this section we study the functional connectivity of the firing-rate
network model introduced in SubSec.~(\ref{sub:The-Firing-Rate-Network-Model}).
For simplicity we consider only the case of two neural populations,
while the theory for an arbitrary number of populations is developed
in the Supplementary Materials. According to Eq.~(\ref{eq:pairwise-covariance}),
the functional connectivity depends on the fundamental matrix of the
network, $\Phi\left(t\right)=e^{\mathcal{J}t}$. In the Supplementary
Materials (see Eq.~(S27)) we calculated $\Phi\left(t\right)$ in
terms of the eigenvalues of the Jacobian matrix $\mathcal{J}$:

\begin{spacing}{0.8}
\begin{center}
{\small{}
\begin{equation}
\lambda_{E}=-\left[\frac{1}{\tau_{E}}+\frac{J_{EE}}{M_{E}}\mathscr{A}_{E}'\left(\mu_{E}\right)\right],\quad\lambda_{I}=-\left[\frac{1}{\tau_{I}}+\frac{J_{II}}{M_{I}}\mathscr{A}_{I}'\left(\mu_{I}\right)\right],\quad\lambda_{0,1}^{\mathcal{R}}=\frac{\mathcal{Y}+\mathcal{Z}\pm\sqrt{\left(\mathcal{Y}-\mathcal{Z}\right)^{2}+4\mathcal{X}}}{2},\label{q:two-populations-eigenvalues}
\end{equation}
}
\par\end{center}{\small \par}
\end{spacing}

\noindent where:

\begin{spacing}{0.8}
\begin{center}
{\small{}
\begin{equation}
\mathcal{X}=\frac{N_{E}N_{I}}{M_{E}M_{I}}J_{EI}J_{IE}\mathscr{A}_{E}'\left(\mu_{E}\right)\mathscr{A}_{I}'\left(\mu_{I}\right),\quad\mathcal{Y}=-\frac{1}{\tau_{E}}+\frac{N_{E}-1}{M_{E}}J_{EE}\mathscr{A}_{E}'\left(\mu_{E}\right),\quad\mathcal{Z}=-\frac{1}{\tau_{I}}+\frac{N_{I}-1}{M_{I}}J_{II}\mathscr{A}_{I}'\left(\mu_{I}\right),\label{eq:parameters-for-the-eigenvalues}
\end{equation}
}
\par\end{center}{\small \par}
\end{spacing}

\noindent and in terms of the functions:

\begin{spacing}{0.8}
\begin{center}
{\small{}
\begin{equation}
K_{\alpha}\overset{\mathrm{def}}{=}\frac{M_{E}\left(\lambda_{\alpha}^{\mathcal{R}}+\frac{1}{\tau_{E}}\right)-\left(N_{E}-1\right)J_{EE}\mathscr{A}_{E}'\left(\mu_{E}\right)}{N_{I}J_{EI}\mathscr{A}_{I}'\left(\mu_{I}\right)}=\frac{N_{E}J_{IE}\mathscr{A}_{E}'\left(\mu_{E}\right)}{M_{I}\left(\lambda_{\alpha}^{\mathcal{R}}+\frac{1}{\tau_{I}}\right)-\left(N_{I}-1\right)J_{II}\mathscr{A}_{I}'\left(\mu_{I}\right)},\quad\alpha=0,1.\label{eq:definition-of-the-terms-K}
\end{equation}
}
\par\end{center}{\small \par}
\end{spacing}

\noindent While we switched to the new notation $\alpha=E,I$, we
keep using $\alpha=0,1$ for the eigenvalues $\lambda_{\alpha}^{\mathcal{R}}$
(and the functions $K_{\alpha}$) because, differently from $\lambda_{E,I}$,
they depend on the parameters of both the populations. If the parameters
of the network are such that $\lambda_{E,I}$ and $\lambda_{0,1}^{\mathcal{R}}$
have negative real part, by applying Eqs.~(\ref{eq:pairwise-covariance})
+ (S27) we end up with the following formula of the covariance matrix
of the membrane potentials $\Sigma^{V}\overset{\mathrm{def}}{=}\underset{t\rightarrow+\infty}{\lim}\left[\mathrm{Cov}\left(V_{i}\left(t\right),V_{j}\left(t\right)\right)\right]_{\forall i,j}$:

\begin{spacing}{0.8}
\begin{center}
{\small{}
\begin{equation}
\Sigma^{V}=\left[\begin{array}{cc}
\varSigma_{EE}^{V} & \varSigma_{EI}^{V}\\
\left[\varSigma{}_{EI}^{V}\right]^{T} & \varSigma_{II}^{V}
\end{array}\right],\label{eq:covariance-matrix-0}
\end{equation}
}
\par\end{center}{\small \par}
\end{spacing}

\noindent \begin{flushleft}
where $T$ denotes the transpose of a matrix, while the blocks $\varSigma_{\alpha\beta}^{V}$
are given by the following formulas:
\par\end{flushleft}

\begin{spacing}{0.8}
\begin{center}
{\footnotesize{}
\begin{align}
\varSigma_{\alpha\alpha}^{V}= & \begin{array}{ccc}
\left(\sigma_{\alpha}^{V}\right)^{2}\mathrm{Id}{}_{N_{\alpha}}+\varsigma_{\alpha\alpha}^{V}\left(\mathbb{I}_{N_{\alpha}}-\mathrm{Id}{}_{N_{\alpha}}\right), &  & \varSigma_{EI}^{V}=\varsigma_{EI}^{V}\mathbb{I}_{N_{E},N_{I}}\end{array}\nonumber \\
\nonumber \\
\left(\sigma_{E}^{V}\right)^{2}= & \left(\sigma_{E}^{\mathscr{B}}\right)^{2}\left\{ \Upsilon_{EE}^{EE}\left[\frac{1}{N_{E}}+C_{EE}^{\mathscr{B}}\left(1-\frac{1}{N_{E}}\right)\right]-\frac{1}{2\lambda_{E}}\left(1-\frac{1}{N_{E}}\right)\left(1-C_{EE}^{\mathscr{B}}\right)\right\} \nonumber \\
 & +\left(\sigma_{I}^{\mathscr{B}}\right)^{2}\Upsilon_{EE}^{II}\left[\frac{1}{N_{I}}+C_{II}^{\mathscr{B}}\left(1-\frac{1}{N_{I}}\right)\right]+2\sigma_{E}^{\mathscr{B}}\sigma_{I}^{\mathscr{B}}\Upsilon_{EE}^{EI}C_{EI}^{\mathscr{B}}\nonumber \\
\nonumber \\
\left(\sigma_{I}^{V}\right)^{2}= & \left(\sigma_{E}^{\mathscr{B}}\right)^{2}\Upsilon_{II}^{EE}\left[\frac{1}{N_{E}}+C_{EE}^{\mathscr{B}}\left(1-\frac{1}{N_{E}}\right)\right]\nonumber \\
 & +\left(\sigma_{I}^{\mathscr{B}}\right)^{2}\left\{ \Upsilon_{II}^{II}\left[\frac{1}{N_{I}}+C_{II}^{\mathscr{B}}\left(1-\frac{1}{N_{I}}\right)\right]-\frac{1}{2\lambda_{I}}\left(1-\frac{1}{N_{I}}\right)\left(1-C_{II}^{\mathscr{B}}\right)\right\} +2\sigma_{E}^{\mathscr{B}}\sigma_{I}^{\mathscr{B}}\Upsilon_{II}^{EI}C_{EI}^{\mathscr{B}}\nonumber \\
\nonumber \\
\varsigma_{EE}^{V}= & \left(\sigma_{E}^{\mathscr{B}}\right)^{2}\left\{ \Upsilon_{EE}^{EE}\left[\frac{1}{N_{E}}+C_{EE}^{\mathscr{B}}\left(1-\frac{1}{N_{E}}\right)\right]+\frac{1}{2\lambda_{E}N_{E}}\left(1-C_{EE}^{\mathscr{B}}\right)\right\} \label{eq:covariance-matrix-1}\\
 & +\left(\sigma_{I}^{\mathscr{B}}\right)^{2}\Upsilon_{EE}^{II}\left[\frac{1}{N_{I}}+C_{II}^{\mathscr{B}}\left(1-\frac{1}{N_{I}}\right)\right]+2\sigma_{E}^{\mathscr{B}}\sigma_{I}^{\mathscr{B}}\Upsilon_{EE}^{EI}C_{EI}^{\mathscr{B}}\nonumber \\
\nonumber \\
\varsigma_{II}^{V}= & \left(\sigma_{E}^{\mathscr{B}}\right)^{2}\Upsilon_{II}^{EE}\left[\frac{1}{N_{E}}+C_{EE}^{\mathscr{B}}\left(1-\frac{1}{N_{E}}\right)\right]\nonumber \\
 & +\left(\sigma_{I}^{\mathscr{B}}\right)^{2}\left\{ \Upsilon_{II}^{II}\left[\frac{1}{N_{I}}+C_{II}^{\mathscr{B}}\left(1-\frac{1}{N_{I}}\right)\right]+\frac{1}{2\lambda_{I}N_{I}}\left(1-C_{II}^{\mathscr{B}}\right)\right\} +2\sigma_{E}^{\mathscr{B}}\sigma_{I}^{\mathscr{B}}\Upsilon_{II}^{EI}C_{EI}^{\mathscr{B}}\nonumber \\
\nonumber \\
\varsigma_{EI}^{V}= & \left(\sigma_{E}^{\mathscr{B}}\right)^{2}\Upsilon_{EI}^{EE}\left[\frac{1}{N_{E}}+C_{EE}^{\mathscr{B}}\left(1-\frac{1}{N_{E}}\right)\right]\nonumber \\
 & +\left(\sigma_{I}^{\mathscr{B}}\right)^{2}\Upsilon_{EI}^{II}\left[\frac{1}{N_{I}}+C_{II}^{\mathscr{B}}\left(1-\frac{1}{N_{I}}\right)\right]+\sigma_{E}^{\mathscr{B}}\sigma_{I}^{\mathscr{B}}\Upsilon_{EI}^{EI}C_{EI}^{\mathscr{B}}.\nonumber 
\end{align}
}
\par\end{center}{\footnotesize \par}
\end{spacing}

\noindent The functions $\Upsilon$ are defined as below:

\begin{spacing}{0.8}
\begin{center}
{\footnotesize{}
\begin{align}
 & \Upsilon_{EE}^{EE}=\frac{1}{\left(K_{1}-K_{0}\right)^{2}}\left[\frac{2K_{0}K_{1}}{\lambda_{0}^{\mathcal{R}}+\lambda_{1}^{\mathcal{R}}}-\frac{1}{2}\left(\frac{K_{0}^{2}}{\lambda_{1}^{\mathcal{R}}}+\frac{K_{1}^{2}}{\lambda_{0}^{\mathcal{R}}}\right)\right],\quad\Upsilon_{EE}^{II}=\frac{1}{\left(K_{1}-K_{0}\right)^{2}}\left[\frac{2}{\lambda_{0}^{\mathcal{R}}+\lambda_{1}^{\mathcal{R}}}-\frac{1}{2}\left(\frac{1}{\lambda_{0}^{\mathcal{R}}}+\frac{1}{\lambda_{1}^{\mathcal{R}}}\right)\right],\nonumber \\
\nonumber \\
 & \Upsilon_{II}^{EE}=\frac{K_{0}^{2}K_{1}^{2}}{\left(K_{1}-K_{0}\right)^{2}}\left[\frac{2}{\lambda_{0}^{\mathcal{R}}+\lambda_{1}^{\mathcal{R}}}-\frac{1}{2}\left(\frac{1}{\lambda_{0}^{\mathcal{R}}}+\frac{1}{\lambda_{1}^{\mathcal{R}}}\right)\right],\quad\Upsilon_{II}^{II}=\frac{1}{\left(K_{1}-K_{0}\right)^{2}}\left[\frac{2K_{0}K_{1}}{\lambda_{0}^{\mathcal{R}}+\lambda_{1}^{\mathcal{R}}}-\frac{1}{2}\left(\frac{K_{0}^{2}}{\lambda_{0}^{\mathcal{R}}}+\frac{K_{1}^{2}}{\lambda_{1}^{\mathcal{R}}}\right)\right],\nonumber \\
\nonumber \\
 & \Upsilon_{EE}^{EI}=\frac{1}{\left(K_{1}-K_{0}\right)^{2}}\left[\frac{1}{2}\left(\frac{K_{0}}{\lambda_{1}^{\mathcal{R}}}+\frac{K_{1}}{\lambda_{0}^{\mathcal{R}}}\right)-\frac{K_{0}+K_{1}}{\lambda_{0}^{\mathcal{R}}+\lambda_{1}^{\mathcal{R}}}\right],\quad\Upsilon_{II}^{EI}=\frac{K_{0}K_{1}}{\left(K_{1}-K_{0}\right)^{2}}\left[\frac{1}{2}\left(\frac{K_{0}}{\lambda_{0}^{\mathcal{R}}}+\frac{K_{1}}{\lambda_{1}^{\mathcal{R}}}\right)-\frac{K_{0}+K_{1}}{\lambda_{0}^{\mathcal{R}}+\lambda_{1}^{\mathcal{R}}}\right],\nonumber \\
\nonumber \\
 & \Upsilon_{EI}^{EE}=\frac{K_{0}K_{1}}{\left(K_{1}-K_{0}\right)^{2}}\left[\frac{K_{0}+K_{1}}{\lambda_{0}^{\mathcal{R}}+\lambda_{1}^{\mathcal{R}}}-\frac{1}{2}\left(\frac{K_{0}}{\lambda_{1}^{\mathcal{R}}}+\frac{K_{1}}{\lambda_{0}^{\mathcal{R}}}\right)\right],\quad\Upsilon_{EI}^{II}=\frac{1}{\left(K_{1}-K_{0}\right)^{2}}\left[\frac{K_{0}+K_{1}}{\lambda_{0}^{\mathcal{R}}+\lambda_{1}^{\mathcal{R}}}-\frac{1}{2}\left(\frac{K_{0}}{\lambda_{0}^{\mathcal{R}}}+\frac{K_{1}}{\lambda_{1}^{\mathcal{R}}}\right)\right],\nonumber \\
\nonumber \\
 & \Upsilon_{EI}^{EI}=\frac{1}{\left(K_{1}-K_{0}\right)^{2}}\left[K_{0}K_{1}\left(\frac{1}{\lambda_{1}^{\mathcal{R}}}+\frac{1}{\lambda_{0}^{\mathcal{R}}}-\frac{2}{\lambda_{0}^{\mathcal{R}}+\lambda_{1}^{\mathcal{R}}}\right)-\frac{K_{0}^{2}+K_{1}^{2}}{\lambda_{0}^{\mathcal{R}}+\lambda_{1}^{\mathcal{R}}}\right].\label{eq:covariance-matrix-2}
\end{align}
}
\par\end{center}{\footnotesize \par}
\end{spacing}

\noindent $\sigma_{\alpha}^{V}$ is the standard deviation of the
neurons in the population $\alpha$, while $\varsigma_{\alpha\alpha}^{V}$
is the covariance between any pair of neurons in the same population
$\alpha$, and $\varsigma_{EI}^{V}$ is the covariance between any
pair of neurons in two different populations.

Finally, by replacing these results into the formula $C_{\alpha\beta}^{V}\overset{\mathrm{def}}{=}\frac{\varsigma_{\alpha\beta}^{V}}{\sigma_{\alpha}^{V}\sigma_{\beta}^{V}}$
obtained from Eq.~(\ref{eq:pairwise-correlation}), we get an expression
of the functional connectivity $C^{V}=\left[C_{\alpha\beta}^{V}\right]_{\forall\left(\alpha,\beta\right)}$
of the network. From this formula we will prove that, depending on
the values of the parameters, the network can switch from local chaos
to critical slowing down. These special regimes have important and
contrasting properties, which will be discussed in detail in SubSecs.~(\ref{sub:Local-Chaos})
and (\ref{sub:Critical-Slowing-Down}).

\subsection{Local Chaos \label{sub:Local-Chaos}}

Local chaos is the condition that characterizes asynchronous neural
states. Its most important features are small amplitude temporal fluctuations
of the membrane potentials, as well as weak cross-correlations between
them. Local chaos can be generated in two different ways. Probably
the most known is the increase of the network's size \cite{Samuelides2007,Touboul2012,Baladron2012a,Baladron2012b}.
Indeed, for $\mathfrak{P}=2$, if $C_{EE}^{\mathscr{B}}=C_{II}^{\mathscr{B}}=C_{EI}^{\mathscr{B}}=0$,
from Eqs.~(\ref{eq:covariance-matrix-1}) + (\ref{eq:covariance-matrix-2})
we observe that $\varsigma_{\alpha\beta}^{V}\rightarrow0$ and $\left(\sigma_{\alpha}^{V}\right)^{2}\rightarrow\left(-\frac{1}{2\lambda_{\alpha}}\right)\left(\sigma_{\alpha}^{\mathscr{B}}\right)^{2}\approx\frac{1}{2\tau_{\alpha}}\left(\sigma_{\alpha}^{\mathscr{B}}\right)^{2}$
(for $\alpha,\beta\in\left\{ E,I\right\} $) in the thermodynamic
limit $N_{E,I}\rightarrow\infty$. In other words, in infinite-size
networks with independent Brownian motions, the membrane potentials
are independent too, leading to local chaos. Local chaos is usually
invoked to justify the mean-field description of large neural networks
and is compatible with recent findings in visual cortex \cite{Ecker2010,Renart2010,Tetzlaff2012}.

\begin{figure}
\begin{centering}
\includegraphics[scale=0.21]{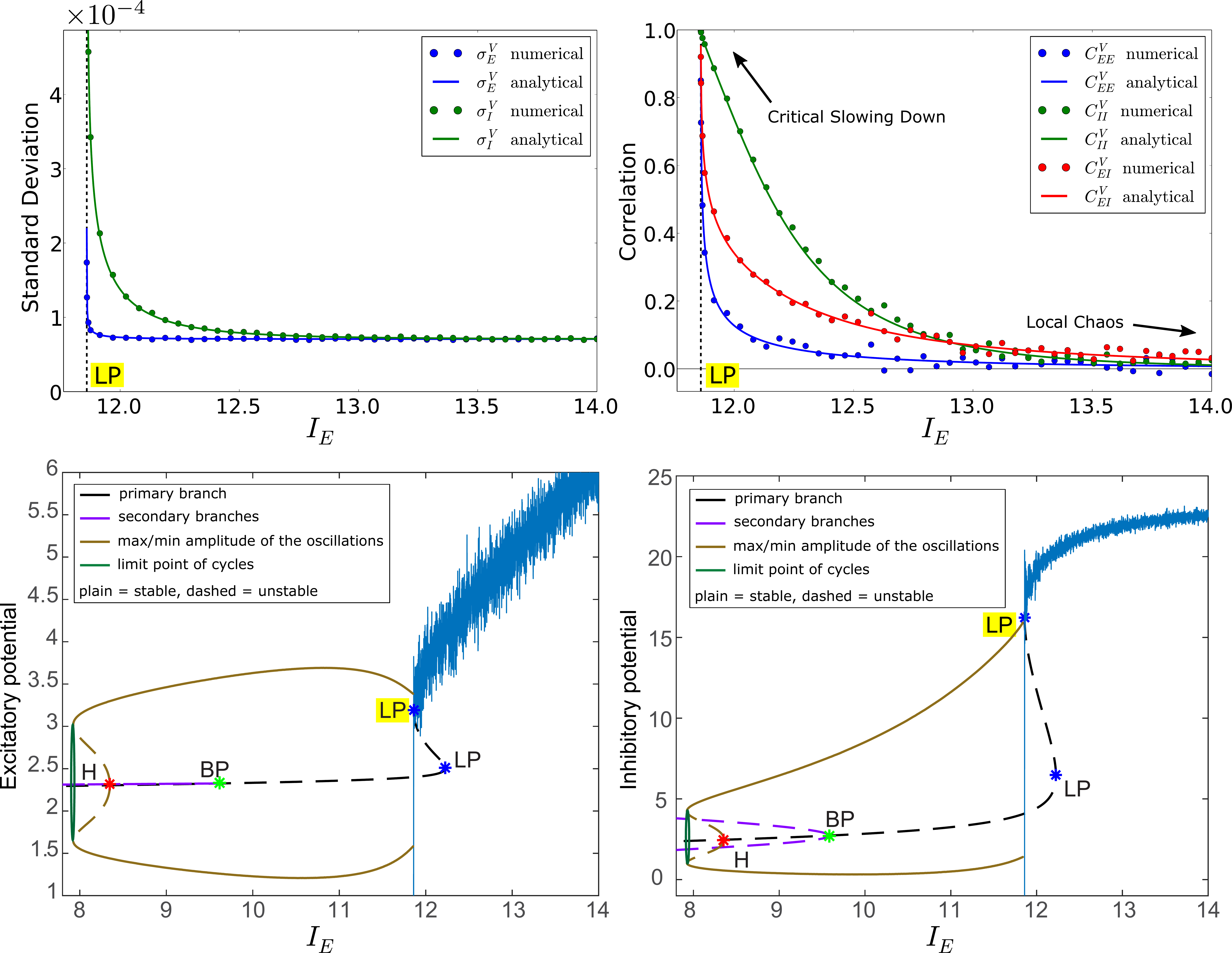}
\par\end{centering}

\protect\caption{\label{fig:local-chaos-and-saddle-node-bifurcation} \small \textbf{Transition
between local chaos and critical slowing down near a saddle-node bifurcation.}
The top panels show a good agreement between the numerical approximations
of the standard deviation and correlation (left and right panel respectively),
and the corresponding analytical formulas (see Eqs.~(\ref{eq:covariance-matrix-1})
+ (\ref{eq:covariance-matrix-2})). The numerical approximations have
been obtained by integrating the neural equations (\ref{eq:exact-rate-equations})
with the Euler-Maruyama method, for the values of the parameters reported
in Tab.~(\ref{tab:network-parameters}) and $I_{I}=-35$. The integration
time step is $\Delta t=0.001$, and the equations are integrated with
a Monte Carlo method over $5,000$ repetitions of the network dynamics
in the temporal interval $t=\left[0,30\right]$. For large inputs
($I_{E}>13$) we observe the formation of local chaos, which is characterized
by weak correlation and low variability. On the other side, near a
saddle-node bifurcation ($I_{E}\approx11.86$, see the highlighted
LP), we observe strong correlations and wide temporal fluctuations
that characterize critical slowing down. The bottom panels show numerical
simulations of the fluctuations of the membrane potentials in the
excitatory and inhibitory population (left and right panel respectively),
calculated at $t=30$ for different values of $I_{E}$ and superposed
to the codimension one bifurcation diagram of the network. The fluctuations
are displayed at $3,000\times$ actual size in the excitatory and
inhibitory population, in order to make them visible on the bifurcation
diagrams. The reader may verify the agreement between the standard
deviations (top-left panel) and the envelope of the fluctuations of
the membrane potentials.}
\end{figure}

Interestingly, also finite-size networks can experience decorrelated
activity. In \cite{Fasoli2015} the authors showed that, for any $N$,
weak correlations occur for strongly depolarizing or strongly hyperpolarizing
external inputs, if the Brownian motions are independent. This phenomenon
can be proven for the two-populations case as a consequence of $\Upsilon_{\alpha\alpha}^{\alpha\alpha}\rightarrow-\frac{1}{2\lambda_{\alpha}}$,
$\Upsilon_{\beta\beta}^{\alpha\alpha}\rightarrow0$ (with $\alpha\neq\beta$)
and $\Upsilon_{EI}^{\alpha\alpha}\rightarrow0$ for $\left|I_{E,I}\right|\rightarrow\infty$,
which in turn is due to the saturation of the activation function
$\mathscr{A}\left(V\right)$. For the same reason, the standard deviations
$\sigma_{E,I}^{V}$ of the neural activity in the two populations
decrease with the input. Interestingly, the reduction of both the
correlation and the variance of the neural responses is supported
by experimental evidence \cite{Tan2014,Ponce-Alvarez2015}. Fig.~(\ref{fig:local-chaos-and-saddle-node-bifurcation})
shows an example of formation of local chaos in a finite-size network
for $\mathfrak{P}=2$, which is obtained for the values of the parameters
in Tab.~(\ref{tab:network-parameters}) and for strong stimuli ($I_{E}>13$,
$I_{I}=-35$). 

We observe that local chaos always requires independent Brownian motions
to occur. However, the theory developed in this article can also be
applied to networks with correlated noise. In \cite{Fasoli2015} the
authors observed that if the Brownian motions are correlated, local
chaos does not occur anymore, neither in large network nor for strong
stimuli. The same result is obtained from Eqs.~(\ref{eq:covariance-matrix-1})
+ (\ref{eq:covariance-matrix-2}). In particular, since $\Upsilon_{\alpha\alpha}^{EI}\rightarrow0$
and $\Upsilon_{EI}^{EI}\rightarrow\frac{1}{2\sqrt{\lambda_{E}\lambda_{I}}}$
for $\left|I_{E,I}\right|\rightarrow\infty$, we get $\left(\sigma_{\alpha}^{V}\right)^{2}\rightarrow\frac{1}{2\tau_{\alpha}}\left(\sigma_{\alpha}^{\mathscr{B}}\right)^{2}$
and $C_{\alpha\beta}^{V}\rightarrow C_{\alpha\beta}^{\mathscr{B}}$
(see also Fig.~(\ref{fig:functional-connectivity-for-correlated-Brownian-motions}),
where we plot the functional connectivity for different values of
$C^{\mathscr{B}}$). In other words, for strong stimuli the correlation
between the membrane potentials converges to that between the Brownian
motions.

\subsection{Critical Slowing Down \label{sub:Critical-Slowing-Down}}

Critical slowing down is the condition that characterizes synchronous
neural states. Contrary to local chaos, its most important features
are large temporal fluctuations of the membrane potentials and strong
cross-correlations. Critical slowing down typically occurs at the
bifurcation points of the system. In \cite{Fasoli2016} we performed
a detailed bifurcation analysis in the case $\mathfrak{P}=2$ and
for the values of the parameters in Tab.~(\ref{tab:network-parameters}),
obtaining the entangled set of local and global bifurcations shown
in Fig.~(\ref{Fig:codimension-two-bifurcation-diagram}). Local bifurcations
occur when a parameter variation causes the stability of an equilibrium
point to change, therefore they are studied through the eigenvalues
of the Jacobian matrix. Local bifurcations can be of codimension one
or two, depending on the number of parameters (i.e. $I_{E,I}$) that
must be changed for the bifurcation to occur. As shown in Fig.~(\ref{Fig:codimension-two-bifurcation-diagram}),
the local bifurcations of codimension one the network undergoes are
saddle-node, Andronov-Hopf and branching-point bifurcations, while
those of codimension two are cusp, Bogdanov-Takens, generalized Hopf
and zero-Hopf bifurcations. The remaining bifurcations are global,
which means they cannot be studied through a local analysis in terms
of the equilibrium points, but rather they require the analysis of
(a part of) the phase portrait of the network. In particular, the
homoclinic, limit point of cycles, and torus \footnote{More precisely, the torus bifurcation is a local bifurcation of the
Poincaré map of a limit cycle of the network \cite{Kuznetsov1998}.
For this reason the torus bifurcation corresponds to a change of stability
of the fixed points of the Poincaré map, and not to a change of stability
of the equilibrium points of Eq.~(\ref{eq:exact-rate-equations}).
For this reason the torus bifurcation cannot be studied through the
eigenvalues of the Jacobian matrix of the network.} curves, are global bifurcations of codimension one, while saddle-node
on invariant circle curves represent the only global bifurcations
of codimension two. As discussed in \cite{Fasoli2016}, for $N_{I}>2$
other kinds of local bifurcations of codimension two and global bifurcations
may occur due to spontaneous-symmetry breaking. Nevertheless, for
simplicity, in this article we restrict our discussion to the case
$N_{I}=2$ when $\mathfrak{P}=2$.

Similarly to \cite{Kuehn2013}, we study the behavior of the functional
connectivity only at the local bifurcations of the network, and in
particular we consider only those of codimension one. These bifurcations
are studied in SubSecs.~(\ref{sub:Saddle-Node-Bifurcations}), (\ref{sub:Andronov-Hopf-Bifurcations}),
(\ref{sub:Branching-Point-Bifurcations}) for the case $\mathfrak{P}=2$
and in Sec.~(S6) of the Supplementary Materials for the case of an
arbitrary $\mathfrak{P}$. Our theory can also be used to study the
behavior of the functional connectivity near local bifurcations of
codimension two, but due to the high variety of the bifurcations the
system exhibits, a complete study is beyond the purpose of this article. 

Finally, to our knowledge no analytical method is known for studying
the global bifurcations of Eq.~(\ref{eq:exact-rate-equations}).
Currently the correlation at the global bifurcations can be studied
only numerically, but this analysis again is beyond the purpose of
the article.

\subsubsection{Saddle-Node Bifurcations \label{sub:Saddle-Node-Bifurcations}}

Saddle-node bifurcations are tipping points at which tiny perturbations
can cause an abrupt and discontinuous change of the equilibrium point
of the system. These bifurcations have been proposed to occur in a
set of dynamical systems such as ocean-climate systems, financial
markets, ecosystems etc \cite{Scheffer2009}. In neuroscience, phenomena
compatible with the presence of saddle-node bifurcations in the cortex
have been observed for example during anesthetic administration at
the edge between conscious and unconscious states \cite{Steyn-Ross2004}.

In \cite{Fasoli2016} we proved that, in the case $\mathfrak{P}=2$,
the network undergoes a saddle-node bifurcation whenever one of the
eigenvalues $\lambda_{0,1}^{\mathcal{R}}$ in Eq.~(\ref{q:two-populations-eigenvalues})
tends to zero. The saddle-node bifurcations are described by the blue
curves in Fig.~(\ref{Fig:codimension-two-bifurcation-diagram}).
In \cite{Fasoli2016} we also proved that a necessary condition for
the formation of these bifurcations is:

\begin{spacing}{0.8}
\begin{center}
{\small{}
\begin{equation}
\frac{N_{E}-1}{N-1}J_{EE}\frac{\nu_{E}^{\mathrm{max}}\Lambda_{E}}{4}\tau_{E}>1,\label{eq:saddle-node-condition}
\end{equation}
}
\par\end{center}{\small \par}
\end{spacing}

\noindent or in other words sufficiently strong self-excitatory weights
are required. From Eq.~(\ref{eq:covariance-matrix-2}) we observe
that for $\lambda_{0}^{\mathcal{R}}\rightarrow0^{-}$ or $\lambda_{1}^{\mathcal{R}}\rightarrow0^{-}$
the functions $\Upsilon$ diverge, therefore the terms proportional
to $\frac{1}{2\lambda_{E,I}}$ in Eq.~(\ref{eq:covariance-matrix-1})
become negligible. This implies $\varsigma_{\alpha\alpha}^{V}\sim\left(\sigma_{\alpha}^{V}\right)^{2}\rightarrow\infty$
and $\varsigma_{EI}^{V}\sim\sigma_{E}^{V}\sigma_{I}^{V}$, therefore
$C_{\alpha\beta}^{V}\sim1$ between every population. Thus, when the
network is close to a saddle-node bifurcation, we observe the emergence
of critical slowing down. Moreover, we obtain a simple relation between
the variances of the two neural populations, namely $\sigma_{I}^{V}\sim K\sigma_{E}^{V}$,
where $K\overset{\mathrm{def}}{=}\underset{\lambda_{0}^{\mathcal{R}}\rightarrow0^{-}}{\lim}K_{0}=\underset{\lambda_{1}^{\mathcal{R}}\rightarrow0^{-}}{\lim}K_{1}=\frac{N_{E}J_{IE}\mathscr{A}_{E}'\left(\mu_{E}\right)}{\frac{M_{I}}{\tau_{I}}-\left(N_{I}-1\right)J_{II}\mathscr{A}_{I}'\left(\mu_{I}\right)}$.
The reader can also verify that $\varsigma_{EI}^{V}>0$ for $\lambda_{\alpha}^{\mathcal{R}}\rightarrow0$
as a consequence of $K>0$, which in turn is due to $J_{IE}>0$ and
$J_{II}<0$. An example of critical slowing down obtained for $I_{E}\approx11.86$,
$I_{I}=-35$ and $C_{EE}^{\mathscr{B}}=C_{II}^{\mathscr{B}}=C_{EI}^{\mathscr{B}}=0$
is reported in Fig.~(\ref{fig:local-chaos-and-saddle-node-bifurcation}).
We observe that this phenomenon occurs even if there is no correlation
between the Brownian motions (i.e. $C_{\alpha\beta}^{\mathscr{B}}=0$),
therefore it is entirely a consequence of the neural interactions
mediated by the synaptic connections.

To conclude, from Eqs.~(\ref{eq:correlation-between-firing-rates})
+ (\ref{eq:correlation-between-neural-activities}) we observe that
$\mathrm{Corr}\left(V_{i}\left(t\right),V_{j}\left(t\right)\right)\rightarrow1$
for every pair of neurons implies $\mathrm{Corr}\left(a_{\mathcal{G}}\left(t\right),a_{\mathcal{H}}\left(t\right)\right)\rightarrow1$.
Therefore the populations become functionally connected also in terms
of their neural activities.

\subsubsection{Andronov-Hopf Bifurcations \label{sub:Andronov-Hopf-Bifurcations}}

Andronov-Hopf bifurcations correspond to the emergence of neural oscillations,
which are thought to play a key role in many cognitive processes \cite{Ward2003}.
In \cite{Fasoli2016} we proved that, in the case $\mathfrak{P}=2$,
the network undergoes an Andronov-Hopf bifurcation whenever $\lambda_{0,1}^{\mathcal{R}}$
in Eq.~(\ref{q:two-populations-eigenvalues}) are complex-conjugate
purely imaginary. The Andronov-Hopf bifurcations are described by
the red curves in Fig.~(\ref{Fig:codimension-two-bifurcation-diagram}).
In \cite{Fasoli2016} we also proved that a necessary condition for
the formation of these bifurcations is:

\begin{spacing}{0.8}
\begin{center}
{\small{}
\begin{equation}
\frac{\nu_{E}^{\mathrm{max}}\Lambda_{E}}{4\mathfrak{z}}>1\quad\mathrm{and}\quad\mathfrak{b}^{2}-4\mathfrak{a}\mathfrak{c}>0\label{eq:Hopf-condition}
\end{equation}
}
\par\end{center}{\small \par}
\end{spacing}

\noindent where:

\begin{spacing}{0.8}
\begin{center}
{\small{}
\begin{align*}
\mathfrak{z}= & \frac{-\mathfrak{b}-\sqrt{\mathfrak{b}^{2}-4\mathfrak{a}\mathfrak{c}}}{2\mathfrak{a}}\\
\\
\mathfrak{a}= & \left(\frac{N_{E}-1}{N-1}J_{EE}\right)^{2}-\frac{N_{E}N_{I}\left(N_{E}-1\right)}{\left(N-1\right)^{2}\left(N_{I}-1\right)}\frac{J_{EE}J_{EI}J_{IE}}{J_{II}}\\
\\
\mathfrak{b}= & -\frac{2}{\tau_{E}}\frac{N_{E}-1}{N-1}J_{EE}+\frac{N_{E}N_{I}}{\left(N-1\right)\left(N_{I}-1\right)}\frac{J_{EI}J_{IE}}{J_{II}}\left(\frac{1}{\tau_{E}}+\frac{1}{\tau_{I}}\right)\\
\\
\mathfrak{c}= & \frac{1}{\tau_{E}^{2}}
\end{align*}
}
\par\end{center}{\small \par}
\end{spacing}

\begin{figure}
\begin{centering}
\includegraphics[scale=0.21]{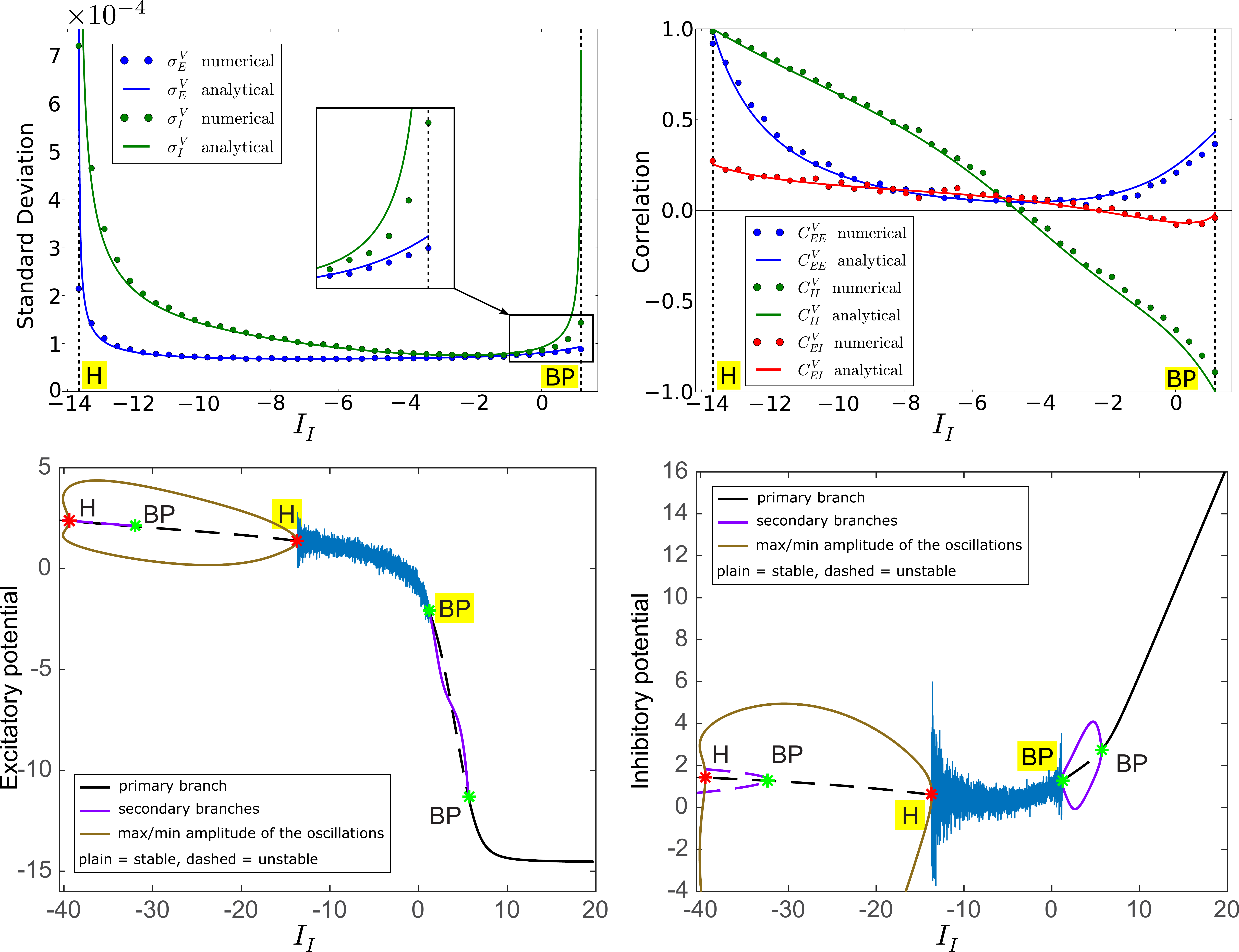}
\par\end{centering}

\protect\caption{\label{fig:Andronov-Hopf-and-branching-point-bifurcations} \small
\textbf{Fluctuations and cross-correlations of the membrane potentials
between Andronov-Hopf and branching-point bifurcations.} The simulations
are similar to those of Fig.~(\ref{fig:local-chaos-and-saddle-node-bifurcation}),
but now we set $I_{E}=1$ and we vary the input to the inhibitory
population, obtaining a transition between an Andronov-Hopf bifurcation
($I_{I}\approx-13.67$, see the highlighted H) and a branching-point
bifurcation ($I_{I}\approx1.165$, highlighted BP). We obtain a good
agreement between numerical and analytical correlations for any current
$I_{I}$ in the range, while the standard deviations display a good
agreement only when $I_{I}$ is sufficiently far from the bifurcation
points. At the Andronov-Hopf and branching-point bifurcations the
standard deviations predicted by the analytical formulas are larger
than those obtained numerically. This suggests that generally second-order
corrections to Eqs.~(\ref{eq:covariance-matrix-1}) + (\ref{eq:covariance-matrix-2})
play a stronger role when the network undergoes a local bifurcation.
Nevertheless, the first-order approximation describes qualitatively
the increase of the standard deviation that characterizes critical
slowing down.}
\end{figure}

\noindent Whenever the network approaches an Andronov-Hopf bifurcation,
we get $\lambda_{0}^{\mathcal{R}}+\lambda_{1}^{\mathcal{R}}\rightarrow0^{-}$,
which causes the terms $\Upsilon$ to diverge (see Eq.~(\ref{eq:covariance-matrix-2})).
For this reason the variance of the membrane potentials diverges as
well, while the cross-correlation tends to one, similarly to the case
of the saddle-node bifurcations. This proves that the network undergoes
critical slowing down also at the Andronov-Hopf bifurcations. An example
obtained for $I_{E}=1$, $I_{I}\approx-13.67$ and $C_{EE}^{\mathscr{B}}=C_{II}^{\mathscr{B}}=C_{EI}^{\mathscr{B}}=0$
is shown in Fig.~(\ref{fig:Andronov-Hopf-and-branching-point-bifurcations}).

\begin{figure}
\begin{centering}
\includegraphics[scale=0.25]{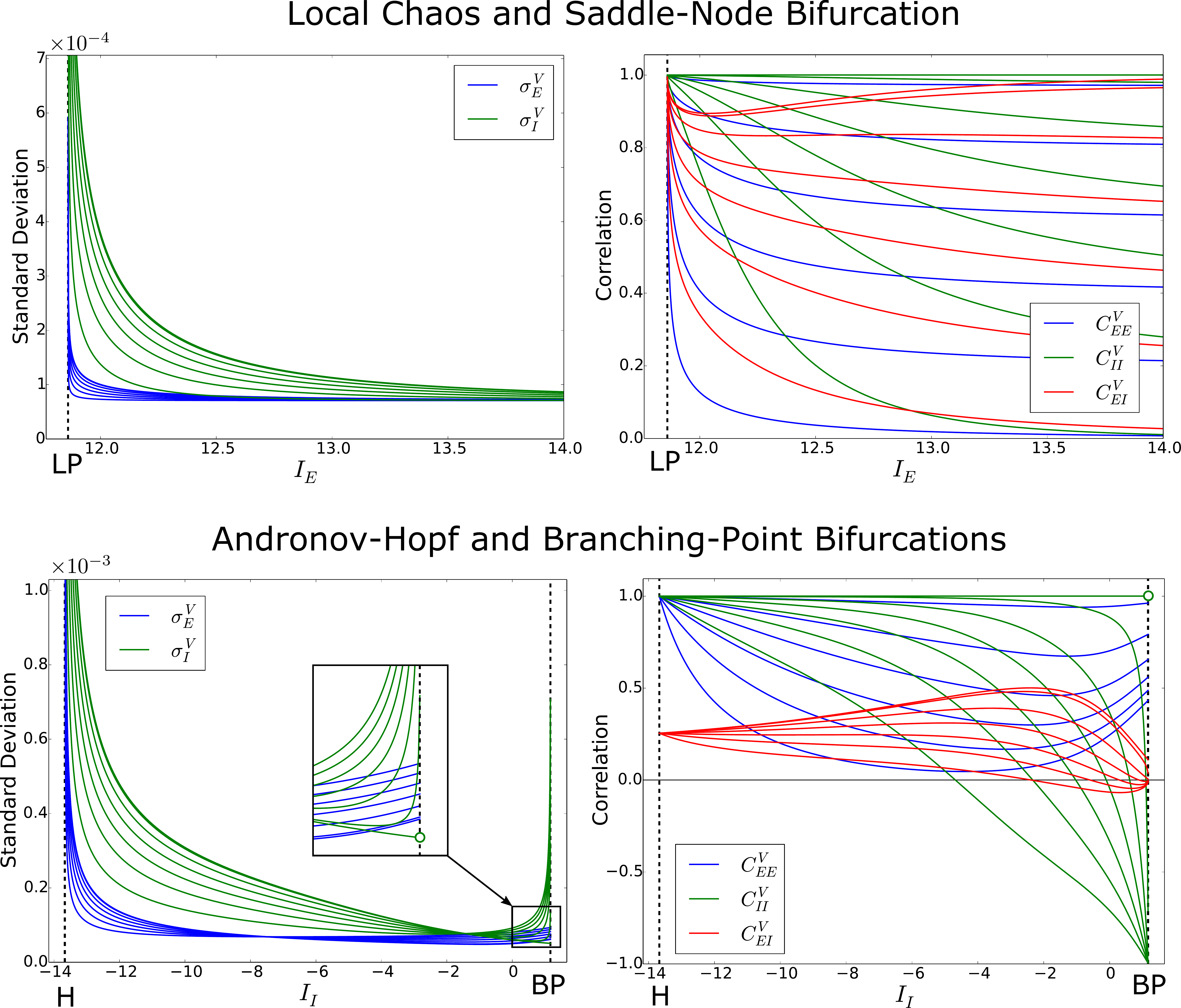}
\par\end{centering}

\protect\caption{\label{fig:functional-connectivity-for-correlated-Brownian-motions}
\small \textbf{Fluctuations and cross-correlations of the membrane
potentials as a function of the input and of the noise correlation.}
The top panels show the standard deviation (left) and the cross-correlation
(right) of the membrane potentials when the network is close to a
saddle-node bifurcation (similarly to Fig.~(\ref{fig:local-chaos-and-saddle-node-bifurcation})),
for different values of the noise correlation. The curves have been
obtained from Eqs.~(\ref{eq:covariance-matrix-1}) + (\ref{eq:covariance-matrix-2})
for $C_{EE}^{\mathscr{B}}=C_{II}^{\mathscr{B}}=C_{EI}^{\mathscr{B}}=0,\,0.2,\,0.4,\,0.6,\,0.8,\,0.97,\,1$.
The panels show that the noise correlation increases both the standard
deviation and the cross-correlation, for $\sigma_{E,I}^{\mathscr{B}}$
fixed (see Tab.~(\ref{tab:network-parameters})). The bottom panels
show similar results for the neural states between Andronov-Hopf and
branching-point bifurcations (compare with Fig.~(\ref{fig:Andronov-Hopf-and-branching-point-bifurcations})).
The only difference is observed close to the branching-point bifurcation,
where $\sigma_{E,I}^{V}$ decrease with the noise correlation.}
\end{figure}

\subsubsection{Branching-Point Bifurcations \label{sub:Branching-Point-Bifurcations}}

In the deterministic model (i.e. for $\sigma_{E,I}^{\mathscr{B}}=0$),
according to the assumptions of Sec.~(\ref{sub:The-Firing-Rate-Network-Model}),
within each population the neurons are dynamically identical, namely
the populations are homogeneous. This means that, in absence of noise,
the network has the symmetry ${\displaystyle \bigtimes_{\alpha=0}^{\mathfrak{P}-1}S_{N_{\alpha}}}$,
where $S_{N_{\alpha}}$ is the permutation group on $N_{\alpha}$
items (also known as \textit{symmetric group}). When we turn on the
noise ($\sigma_{E,I}^{\mathscr{B}}>0$), we introduce a small \textsl{explicit
symmetry-breaking} into Eq.~(\ref{eq:exact-rate-equations}). However,
the behavior of a nearly symmetric dynamical system is more similar
to that of an idealized symmetric system than that of a completely
asymmetric one \cite{Stewart2003}. Therefore it is legitimate to
study Eq.~(\ref{eq:exact-rate-equations}) as a perturbation of the
corresponding deterministic system, if the degree of explicit heterogeneity
introduced by the noise is not too strong. However, symmetry-breaking
may occur also in the deterministic model. Indeed, at the branching-point
bifurcations we observe the formation of a \textit{spontaneous symmetry-breaking}
\cite{Fasoli2016}, because some of the neurons within a given inhibitory
population become dynamically distinct from the others. In other words,
we observe the formation of an heterogeneous inhibitory population,
even if the neural equations (\ref{eq:exact-rate-equations}) for
$\sigma_{E,I}^{\mathscr{B}}=0$ do not contain any term that breaks
explicitly the symmetry. Interestingly, this phenomenon is a consequence
of the finite size of the network, therefore it does not occur in
the thermodynamic limit \cite{Fasoli2016}.

In \cite{Fasoli2016} we also proved that, in the case $\mathfrak{P}=2$,
a branching-point bifurcation occurs whenever $\lambda_{I}=0$ (see
the light green dot-dashed curves in Fig.~(\ref{Fig:codimension-two-bifurcation-diagram}))
and that a necessary condition for their formation is:

\begin{spacing}{0.8}
\begin{center}
{\small{}
\begin{equation}
\frac{\tau_{I}\left|J_{II}\right|\nu_{I}^{\mathrm{max}}\Lambda_{I}}{4\left(N-1\right)}>1.\label{eq:branching-point-condition}
\end{equation}
}
\par\end{center}{\small \par}
\end{spacing}

\noindent This means that sufficiently strong self-inhibitory weights
are required for the bifurcation to occur. According to Eq.~(\ref{eq:covariance-matrix-1}),
for $C_{II}^{\mathscr{B}}<1$ and $\lambda_{I}\rightarrow0^{-}$ only
the variance of the inhibitory neurons diverges. As a consequence,
in the case $C_{II}^{\mathscr{B}}<1$ we get $\left(\sigma_{I}^{V}\right)^{2}\sim\left(\sigma_{I}^{\mathscr{B}}\right)^{2}\left[-\frac{1}{2\lambda_{I}}\left(1-\frac{1}{N_{I}}\right)\left(1-C_{II}^{\mathscr{B}}\right)\right]$
and $\varsigma_{II}^{V}\sim\left(\sigma_{I}^{\mathscr{B}}\right)^{2}\frac{1}{2\lambda_{I}N_{I}}\left(1-C_{II}^{\mathscr{B}}\right)$,
from which we conclude that $C_{II}^{V}\sim\frac{1}{1-N_{I}}$. According
to \cite{Fasoli2015}, this is the lower bound of the correlation
between fully-connected neurons in a homogeneous population with size
$N_{I}$. Since $\frac{1}{1-N_{I}}<0$ for $N_{I}\geq2$, at the branching-point
bifurcations the inhibitory neurons are maximally anti-correlated.
Moreover, according to this formula, correlation tends to $-1$ only
for $N_{I}=2$. Therefore we conclude that, contrary to the saddle-node
and Andronov-Hopf bifurcations, at the branching points critical slowing
down occurs only in the inhibitory population. This result is confirmed
by Fig.~(\ref{fig:Andronov-Hopf-and-branching-point-bifurcations}),
which shows an example obtained for $I_{E}=1$, $I_{I}\approx1.165$
and $C_{EE}^{\mathscr{B}}=C_{II}^{\mathscr{B}}=C_{EI}^{\mathscr{B}}=0$.
Intuitively, the membrane potentials become anti-correlated because
the inhibitory neurons follow different branches of stationary solutions
beyond the branching-point (see the codimension one bifurcation diagram
in the bottom-right panel of Fig.~(\ref{fig:Andronov-Hopf-and-branching-point-bifurcations})).
Therefore while the potential of one neuron increases due to noise
fluctuations, the potential of the other neuron decreases and viceversa,
resulting in a negative correlation.

On the other side, for $C_{II}^{\mathscr{B}}=1$ and $\lambda_{I}\rightarrow0^{-}$,
from Eq.~(\ref{eq:covariance-matrix-1}) we get:

\begin{spacing}{0.8}
\begin{center}
{\small{}
\[
\left(\sigma_{I}^{V}\right)^{2}=\varsigma_{II}^{V}=\left(\sigma_{E}^{\mathscr{B}}\right)^{2}\Upsilon_{II}^{EE}\left[\frac{1}{N_{E}}+C_{EE}^{\mathscr{B}}\left(1-\frac{1}{N_{E}}\right)\right]+\left(\sigma_{I}^{\mathscr{B}}\right)^{2}\Upsilon_{II}^{II}+2\sigma_{E}^{\mathscr{B}}\sigma_{I}^{\mathscr{B}}\Upsilon_{II}^{EI}C_{EI}^{\mathscr{B}},
\]
}
\par\end{center}{\small \par}
\end{spacing}

\noindent therefore now $\left(\sigma_{I}^{V}\right)^{2}$ does not
diverge anymore and $C_{II}^{V}=1$ (see also Fig.~(\ref{fig:functional-connectivity-for-correlated-Brownian-motions})).
To conclude, for $C_{II}^{\mathscr{B}}=1$ and $\lambda_{I}=0$, Eq.~(\ref{eq:covariance-matrix-1})
gives an indeterminate form $\frac{0}{0}$ for the variance $\left(\sigma_{I}^{V}\right)^{2}$,
which is represented by the empty circles in the bottom panels of
Fig.~(\ref{fig:functional-connectivity-for-correlated-Brownian-motions}).
This result can be intuitively interpreted as the consequence of the
competition between the positive correlation introduced by the Brownian
motions and the anti-correlation generated by the branching point.

\section{Discussion \label{sec:Discussion}}

We developed a theory of the functional connectivity of a multi-population
firing-rate network model of arbitrary size and with all-to-all topology.
In particular, this theory can be used for studying the functional
connectivity of small networks composed of a few tens of neurons,
such as cortical microcolumns in mammals \cite{Mountcastle1997} and
neural circuits in some invertebrates \cite{Williams1988}. Our study
relies on the methods introduced in \cite{Fasoli2015,Fasoli2016},
which are not based on statistical averages. For this reason, our
theory can be applied to networks of arbitrary size $N$, and is not
restricted to large networks as in previous works on neural systems
\cite{Ginzburg1994,Bressloff2009,Renart2010,Buice2013}.

The model we introduced is largely analytically tractable, and allowed
us to derive explicit expressions for the functional connectivity
of the network in terms of the cross-correlations between neurons
or neural populations. Then we studied the behavior of the functional
connectivity in terms of the stimuli $I_{\alpha}$, and this analysis
revealed the ability of the network to switch dynamically from asynchronous
regimes characterized by weak correlation and low variability, to
synchronous regimes characterized by strong correlations and wide
temporal fluctuations of the state variables.

The asynchronous regime, known as \textit{local chaos} in the mathematical
literature, can be observed in large networks driven by independent
sources of noise \cite{Samuelides2007,Touboul2012,Baladron2012a,Baladron2012b}.
In this article we proved that local chaos can be generated dynamically
also by strong stimuli in small networks. The decrease of both the
variance and the cross-correlation of the neural responses with the
input occurs only in networks with saturating activation functions,
and it is supported by experimental evidence \cite{Tan2014,Ponce-Alvarez2015}. 

On the other side, the synchronous regime occurs near the bifurcation
points of the network, which are analytically known \cite{Fasoli2016}.
In particular, in the present article we considered the local bifurcations
of codimension one, namely the saddle-node, Andronov-Hopf and branching-point
bifurcations. Contrary to the strongly positive correlations that
occur at the saddle-node and Andronov-Hopf bifurcations, at the branching
points we have observed the emergence of strong \textit{anti-correlations}
between inhibitory neurons.

The emergence of strong correlations at any of the local bifurcations
of the network is a finite-size effect, and does not require correlated
sources of noise. Indeed, for a network with independent Brownian
motions, in \cite{Fasoli2015} the authors proved that the neurons
are strongly synchronized at a time instant $t_{N}$ that depends
on the size of the network. $t_{N}\rightarrow\infty$ in the limit
$N\rightarrow\infty$, therefore strong correlations are very unlikely
to occur in large networks after short time intervals. However, exceptions
may arise in sparsely-connected networks (see SubSec.~(\ref{sub:Future-Directions})),
or if the Brownian motions are correlated.

In SubSec.~(\ref{sub:The-Role-of-Correlations-in-Encoding-and-Integrating-Stimulus-Information})
we explain the importance of the synchronous and asynchronous regimes
for the encoding and integration of stimulus information, while in
SubSec.~(\ref{sub:Spontaneous-Symmetry-Breaking-as-the-Origin-of-Anti-Correlations})
we discuss how spontaneous symmetry-breaking is responsible for negative
correlations. In SubSec.~(\ref{sub:The-Effect-of-Drugs-on-the-Functional-Connectivity})
we explain possible interactions between drugs and the functional
connectivity of the network, and to conclude, in SubSec.~(\ref{sub:Future-Directions})
we discuss future extensions of this work.

\subsection{The Role of Correlations in Encoding and Integrating Stimulus Information
\label{sub:The-Role-of-Correlations-in-Encoding-and-Integrating-Stimulus-Information}}

Local chaos is very undesirable for the sake of functional integration
since it is synonym of functional disconnection and of low information
flow between neurons or populations. Notwithstanding, as explained
in \cite{Abbott1999}, local chaos proves very convenient in the population
encoding of stimulus information. Intuitively, its role can be understood
by observing that:

\begin{spacing}{0.8}
\noindent \begin{center}
{\small{}
\begin{equation}
\mathrm{Var}\left(a_{\mathcal{G}}\left(t\right)\right)=\frac{1}{N_{\mathcal{G}}^{2}}\left[\sum_{i\in\mathcal{G}}\mathrm{Var}\left(\nu_{i}\left(t\right)\right)+{\displaystyle \sum_{\substack{i,j\in\mathcal{G}\\
i\neq j
}
}}\mathrm{Cov}\left(\nu_{i}\left(t\right),\nu_{j}\left(t\right)\right)\right]=\frac{1}{N_{\mathcal{G}}}\mathrm{Var}\left(\mathcal{G}\right)+\frac{N_{\mathcal{G}}-1}{N_{\mathcal{G}}}\mathrm{Cov}\left(\mathcal{G},\mathcal{G}\right),\label{eq:variance-of-the-neural-activity}
\end{equation}
}
\par\end{center}{\small \par}
\end{spacing}

\noindent where we defined $\mathrm{Var}\left(\mathcal{G}\right)\overset{\mathrm{def}}{=}\left.\mathrm{Var}\left(\nu_{i}\left(t\right)\right)\right|_{i\in\mathcal{G}}$
and $\mathrm{Cov}\left(\mathcal{G},\mathcal{G}\right)\overset{\mathrm{def}}{=}\left.\mathrm{Cov}\left(\nu_{i}\left(t\right),\nu_{j}\left(t\right)\right)\right|_{i,j\in\mathcal{G},\,i\neq j}$.
The last equality of Eq.~(\ref{eq:variance-of-the-neural-activity})
holds only if $\mathcal{G}$ is a subset of a neural population where
spontaneous symmetry-breaking did not occur (so that all the neurons
in $\mathcal{G}$ are homogeneous), and it shows that $\underset{N_{\mathcal{G}}\rightarrow\infty}{\lim}\mathrm{Var}\left(a_{\mathcal{G}}\left(t\right)\right)=0$
only if $\underset{N_{\mathcal{G}}\rightarrow\infty}{\lim}\mathrm{Cov}\left(\mathcal{G},\mathcal{G}\right)=0$
and $\mathrm{Var}\left(\mathcal{G}\right)\sim N^{\varphi}$ with $\varphi<1$.
Therefore, the variability of the neural activity of a large population
is much smaller than that of a single neuron only in the local chaos
regime. This means that the population activity could be used to encode
the stimulus information reliably. In synchronous states $\mathrm{Var}\left(a_{\mathcal{G}}\left(t\right)\right)$
does not tend to zero for large $N_{\mathcal{G}}$, since $\underset{N_{\mathcal{G}}\rightarrow\infty}{\lim}\mathrm{Cov}\left(\mathcal{G},\mathcal{G}\right)\neq0$.
This is the reason why strong correlations are commonly thought to
degrade the performance of population encoding. In experiments with
macaques, Ecker et al \cite{Ecker2010} showed that the information
about the stimulus which is conveyed by intra-columnar neurons in
the primary visual cortex is actually increased by local chaos. This
confirms the role of small correlations in improving the encoding
accuracy of large neural populations.

On the other side, for small networks local chaos improves again the
encoding accuracy of the population, but the term $\frac{1}{N_{\mathcal{G}}}\mathrm{Var}\left(\mathcal{G}\right)$
in Eq.~(\ref{eq:variance-of-the-neural-activity}) may still be large
for $\mathrm{Cov}\left(\mathcal{G},\mathcal{G}\right)\rightarrow0$,
depending on the variance of the noise $\left(\sigma_{E,I}^{\mathscr{B}}\right)^{2}$.
For this reason small neural circuits may need to rely on other mechanisms
for encoding information. In \cite{Abbott1999} Abbott and Dayan proved,
somewhat contrary to intuition, that the information about the stimulus
conveyed by the network increases in synchronous states with multiplicative
noise (i.e. stimulus-dependent) statistics, if the neurons have heterogeneous
responses. Due to the non-linearity of the activation function (\ref{eq:algebraic-activation-function}),
the cross-correlation structure of our model depends on the stimulus,
therefore the noise is multiplicative. Moreover, realistic networks
have some degree of heterogeneity in the distribution of their parameters
(for example see \cite{Fasoli2015} for the extension of our results
to networks with heterogeneous synaptic weights). We thus expect that
when a biological network is close to a bifurcation point, despite
the presence of correlations and the increase in the variability of
the neural activity, critical slowing down strongly increases its
encoding accuracy, even if the network is small-sized. This proves
that at the bifurcation points both the encoding and the integration
of stimulus information are enhanced by correlations. In turn, this
result suggests that the bifurcation points may be an ideal place
for the brain to accomplish its functions. This intuition is supported
by several experimental findings, that advance the hypothesis that
self-organization naturally maintains the brain near criticality \cite{Werner2007,Chialvo2010}.

\subsection{Spontaneous Symmetry-Breaking as the Origin of Anti-Correlations
\label{sub:Spontaneous-Symmetry-Breaking-as-the-Origin-of-Anti-Correlations}}

We proved that at the branching-point bifurcations the inhibitory
neurons become strongly anti-correlated as a consequence of spontaneous
symmetry-breaking. More generally, other kinds of spontaneous symmetry-breaking
can occur in the network, depending on its symmetries. For example,
in the case of two identical inhibitory populations, two different
symmetries may be broken: the symmetry between neurons in a given
population, and that between the two populations. In the latter case,
the two populations would behave differently from each other, while
keeping their corresponding neurons homogeneous. This phenomenon is
also characterized by strongly positive intra-population correlations
and strongly negative inter-population correlations (result not shown),
reinforcing the idea of a general relationship between spontaneous
symmetry-breaking and anti-correlations. In \cite{Fasoli2016} we
described possible extensions of our formalism to spatially extended
networks with more complex symmetries, therefore spontaneous symmetry-breaking
is likely to affect also the cross-correlation structure of large-scale
neural models.

Negative correlations have been observed in resting-state fMRI experiments,
for example during cognitive tasks performed by human subjects \cite{Fox2005},
and also in the frontolimbic circuit of awake rats \cite{Liang2012},
but their origin and functional role are still poorly understood.
Our findings suggest branching-point bifurcations and spontaneous
symmetry-breaking as a potential neurobiological basis of this phenomenon.

\subsection{The Effect of Drugs on the Functional Connectivity \label{sub:The-Effect-of-Drugs-on-the-Functional-Connectivity}}

Our model may predict how information encoding and integration in
cortical circuits are affected at the mesoscopic scale by drugs. This
study can be performed through a bifurcation analysis in terms of
the synaptic weights $J_{\alpha\beta}$, in order to simulate the
effect of drugs on the excitability of the neural tissue. Glutamate
is the major excitatory neurotransmitter in the adult brain, and some
drugs such as \textit{memantine} and \textit{lamotrigine} inhibit
its release \cite{Chen1992,Cunningham2000}, resulting in a reduction
of the excitatory weights. On the contrary, other compounds such as
the \textit{ibotenic acid} \cite{Krogsgaard-Larsen1980} are glutamate
receptor agonists, therefore they result in an increase of the excitatory
weights. Furthermore, the primary inhibitory neurotransmitter in the
brain is the $\gamma$-Aminobutyric acid (GABA) and some drugs such
as \textit{bicuculline} and \textit{pentylenetetrazol} reduce its
release \cite{Curtis1971,Corda1992}, while others such as \textit{propofol},
\textit{thiopental} and \textit{isoflurane} enhance it \cite{Garcia2010}.
Therefore their administration causes a reduction or an increase of
the inhibitory synaptic weights respectively. This suggests that by
modifying the parameters $J_{\alpha\beta}$ we may study how drugs
affect the functional connectivity of the cortex at the mesoscopic
scale. In particular, whenever for a set of synaptic weights the network
does not satisfy the conditions (\ref{eq:saddle-node-condition}),
(\ref{eq:Hopf-condition}), (\ref{eq:branching-point-condition}),
the corresponding bifurcations become forbidden for any pair of stimuli
$\left(I_{E},I_{I}\right)$. This means that the network cannot rely
anymore on its local bifurcation points for integrating information.
It is therefore natural to speculate that this phenomenon may provide
a neurobiological basis for the strong cognitive impairments observed
in drug users \cite{Williamson1997}.

\subsection{Future Directions \label{sub:Future-Directions}}

We studied the functional connectivity of multi-population networks
near local bifurcations of codimension one. Furthermore, our theory
can be easily extended to the analysis of local bifurcations of larger
codimension, while the lack of analytical techniques restricts the
study of \textit{global} bifurcations to numerical methods only.

Another possible extension of our theory is the study of the functional
connectivity of sparse networks. In \cite{Fasoli2015} the authors
showed that when the number of connections per neuron does not diverge
for $N\rightarrow\infty$, local chaos in general does not occur in
the thermodynamic limit for weak stimuli. Therefore in sufficiently
sparse networks asynchronous states can be generated only through
strong stimuli. Moreover, in \cite{Fasoli2016} we showed that in
sparse networks the branching-point bifurcations are more likely to
occur, even in excitatory populations, resulting in a considerable
increase of the complexity of the bifurcation diagrams.

To conclude, it is possible to study the functional connectivity of
small neural circuits with random synaptic weights, extending the
results obtained in \cite{Ginzburg1994} for large random networks.
In \cite{Fasoli2015} the authors introduced explicit formulas for
the calculation of the functional connectivity of networks with random
weights, but their bifurcation structure is still unexplored. We started
to tackle the problem in \cite{Fasoli2016}, where we showed that
random synaptic weights cause an explicit symmetry-breaking between
neurons and therefore the removal of the degeneracy of the eigenvalues.
Depending on the degree of heterogeneity, random weights result in
a further increase of the complexity of the bifurcation diagrams,
which we will investigate in future work.

\section*{Acknowledgments}

This research was supported by the Autonomous Province of Trento,
Call ``Grandi Progetti 2012,\textquotedbl{} project ``Characterizing
and improving brain mechanisms of attention\textemdash ATTEND\textquotedbl{},
and by the Future and Emerging Technologies (FET) programme within
the Seventh Framework Programme for Research of the European Commission,
under FET-Open grant FP7-600954 (VISUALISE).

\noindent \begin{flushleft}
The funders had no role in study design, data collection and analysis,
decision to publish, interpretation of results, or preparation of
the manuscript.
\par\end{flushleft}

\bibliographystyle{plain}
\bibliography{Bibliography}

\end{document}


\noindent \textbf{\Large{}From Local Chaos to Critical Slowing Down:
A Theory of the Functional Connectivity of Small Neural Circuits }\\
\textbf{Supplementary Materials}\textbf{\Large{}}\\
{\Large \par}

\noindent \begin{flushleft}
Diego Fasoli$^{1,\ast}$, Anna Cattani$^{1}$, Stefano Panzeri$^{1}$
\par\end{flushleft}

\medskip{}

\noindent \begin{flushleft}
\textbf{{1} Laboratory of Neural Computation, Center for Neuroscience
and Cognitive Systems @UniTn, Istituto Italiano di Tecnologia, 38068
Rovereto, Italy }
\par\end{flushleft}

\noindent \begin{flushleft}
\textbf{$\ast$ Corresponding Author. E-mail: diego.fasoli@iit.it}
\par\end{flushleft}

\section*{Structure of the Supplementary Materials \label{sec:Structure-of-the-Supplementary-Materials}}

The Supplementary Materials are organized as follows. First, in Sec.~\eqref{sec:Jacobian-Matrix}
we derive the formula of the Jacobian matrix $\mathcal{J}$ of the
multi-population network, whose eigenvalues and eigenvectors are calculated
in Sec.~\eqref{sec:Eigenvalues-and-Eigenvectors}. Then, in Sec.~\eqref{sec:Fundamental-Matrix}
we derive the fundamental matrix $\Phi\left(t\right)=e^{\mathcal{J}t}$
of the system. In Sec.~\eqref{sec:Constraints-for-the-Covariance-Matrix-of-the-Noise}
we show the constraints that must be satisfied by the covariance matrix
of the noise $\Sigma^{\mathscr{B}}$ in order to be a genuine covariance
matrix, while in Sec.~\eqref{sec:Functional-Connectivity} we use
all these results to calculate the functional connectivity of the
network in terms of its cross-correlation structure. In particular,
in SubSec.~\eqref{sub:Weak-Correlation-Regime-(Local-Chaos)} we
use the formula of the functional connectivity to prove that the network
undergoes a weak-correlation regime with small fluctuations (local
chaos) when its size is increased to infinity or when the neurons
are stimulated by strong inputs. On the other side, in SubSec. \eqref{sub:Strong-Correlation-Regime-(Local-Bifurcations)}
we prove that when the network approaches a bifurcation point, the
neural activity undergoes a strong-correlation regime with large fluctuations
(critical slowing down). In particular, we discuss only the local
bifurcations of codimension one, namely saddle-node, Andronov-Hopf
and branching-point bifurcations.

\section{Jacobian Matrix \label{sec:Jacobian-Matrix}}

From now on we use the notation $\left[\mathscr{M}_{\alpha\beta}\right]_{\forall\left(\alpha,\beta\right)}$
to represent a block matrix containing $\mathfrak{P}^{2}$ blocks
$\mathscr{M}_{\alpha\beta}$ for $\alpha,\beta=0,\ldots,\mathfrak{P}-1$.
Thus, for example the synaptic connectivity matrix (see Eq.~(3) in
the main text) can be written as $J=\left[\mathfrak{J}_{\alpha\beta}\right]_{\forall\left(\alpha,\beta\right)}$.
We use the same notation also for $\mathfrak{P}\times\mathfrak{P}$
matrices. Thus, for example the reduced Jacobian matrix defined later
in Eq.~\eqref{eq:reduced-Jacobian-matrix} can be written as $\mathcal{J}^{\mathcal{R}}=\left[\mathcal{J}_{\alpha\beta}^{\mathcal{R}}\right]_{\forall\left(\alpha,\beta\right)}$.

From Eq.~(1) in the main text we get that the full Jacobian matrix
of the network is $\mathcal{J}=\left[\mathscr{J}_{\alpha\beta}\right]_{\forall\left(\alpha,\beta\right)}$,
where:

\begin{spacing}{0.8}
\begin{center}
{\small{}
\begin{equation}
\mathscr{J}_{\alpha\beta}=\begin{cases}
-\frac{1}{\tau_{\alpha}}\mathrm{Id}{}_{N_{\alpha}}+\frac{J_{\alpha\alpha}}{M_{\alpha}}\mathscr{A}_{\alpha}'\left(\mu_{\alpha}\right)\left(\mathbb{I}_{N_{\alpha}}-\mathrm{Id}_{N_{\alpha}}\right), & \;\mathrm{for}\;\alpha=\beta\\
\\
\frac{J_{\alpha\beta}}{M_{\alpha}}\mathscr{A}_{\beta}'\left(\mu_{\beta}\right)\mathbb{I}_{N_{\alpha},N_{\beta}}, & \;\mathrm{for}\;\alpha\neq\beta.
\end{cases}\label{eq:Jacobian-matrix-multi-population}
\end{equation}
}
\par\end{center}{\small \par}
\end{spacing}

\noindent $\mu_{\alpha}$ are the stationary membrane potentials obtained
form Eq.~(1) when the network has constant input $I_{\alpha}$ and
the noise is not present ($\sigma_{i}^{\mathscr{B}}=0\;\forall i$).
In other words, $\mu_{\alpha}$ are the solutions of the following
system of non-linear algebraic equations:

\begin{spacing}{0.8}
\begin{center}
{\small{}
\begin{equation}
-\frac{1}{\tau_{\alpha}}\mu_{\alpha}+\frac{N_{\alpha}-1}{M_{\alpha}}J_{\alpha\alpha}\mathscr{A}_{\alpha}\left(\mu_{\alpha}\right)+{\displaystyle \sum_{\substack{\beta=0\\
\beta\neq\alpha
}
}^{\mathfrak{P}-1}}\frac{N_{\beta}}{M_{\alpha}}J_{\alpha\beta}\mathscr{A}_{\beta}\left(\mu_{\beta}\right)+I_{\alpha}=0,\quad\alpha=0,\ldots,\mathfrak{P}-1.\label{eq:equilibrium-points-equations-multi-population}
\end{equation}
}
\par\end{center}{\small \par}
\end{spacing}

\noindent Eq.~\eqref{eq:equilibrium-points-equations-multi-population}
must be solved numerically, or by the asymptotic perturbative expansion
that we introduced in the Supplementary Materials of \cite{Fasoli2016}.

\section{Eigenvalues and Eigenvectors \label{sec:Eigenvalues-and-Eigenvectors}}

For every $\alpha=0,\ldots,\mathfrak{P}-1$, it is trivial to prove
that the Jacobian matrix \eqref{eq:Jacobian-matrix-multi-population}
has $N_{\alpha}-1$ eigenvalues of the form:

\begin{spacing}{0.8}
\begin{center}
{\small{}
\begin{equation}
\lambda_{\alpha}^{\mathcal{P}}=-\left[\frac{1}{\tau_{\alpha}}+\frac{J_{\alpha\alpha}}{M_{\alpha}}\mathscr{A}_{\alpha}'\left(\mu_{\alpha}\right)\right],\label{eq:Jacobian-matrix-degenerate-eigenvalues}
\end{equation}
}
\par\end{center}{\small \par}
\end{spacing}

\noindent with corresponding eigenvectors:

\begin{spacing}{0.8}
\begin{center}
{\footnotesize{}
\begin{equation}
\boldsymbol{v}_{\alpha;i}^{\mathcal{P}}=\left[\begin{array}{c}
0\\
\vdots\\
0\\
1\\
\vdots\\
1\\
1-N_{\alpha}\\
1\\
\vdots\\
1\\
0\\
\vdots\\
0
\end{array}\right],\quad i=0,\ldots,N_{\alpha}-2.\label{eq:Jacobian-matrix-eigenvectors-0}
\end{equation}
}
\par\end{center}{\footnotesize \par}
\end{spacing}

\noindent The term $1-N_{\alpha}$ is the $\left(n_{\alpha-1}+i\right)$th
entry of the vector $\boldsymbol{v}_{\alpha;i}^{\mathcal{P}}$, where:

\begin{spacing}{0.8}
\begin{center}
{\small{}
\begin{equation}
n_{\alpha-1}\overset{\mathrm{def}}{=}{\displaystyle \sum_{p=0}^{\alpha-1}}N_{p}\label{eq:population-index-bordes}
\end{equation}
}
\par\end{center}{\small \par}
\end{spacing}

\noindent with $n_{-1}\overset{\mathrm{def}}{=}0$. In Eq.~\eqref{eq:Jacobian-matrix-eigenvectors-0},
the $\left(n_{\alpha-1}+j\right)$th entries of $\boldsymbol{v}_{\alpha;i}^{\mathcal{P}}$
(with $j=0,\ldots,N_{\alpha}-1$ and $j\neq i$) are equal to $1,$
while all the remaining entries are equal to $0$.

Now we prove that the remaining $\mathfrak{P}$ eigenvalues of $\mathcal{J}$,
that we call $\lambda_{\alpha}^{\mathcal{R}}$, are those of the $\mathfrak{P}\times\mathfrak{P}$
matrix $\mathcal{J}^{\mathcal{R}}=\left[\mathcal{J}_{\alpha\beta}^{\mathcal{R}}\right]_{\forall\left(\alpha,\beta\right)}$
(the ``reduced'' Jacobian matrix of the network), where:

\begin{spacing}{0.8}
\begin{center}
{\small{}
\begin{equation}
\mathcal{J}_{\alpha\beta}^{\mathcal{R}}=\begin{cases}
-\frac{1}{\tau_{\alpha}}+\frac{N_{\alpha}-1}{M_{\alpha}}J_{\alpha\alpha}\mathscr{A}_{\alpha}'\left(\mu_{\alpha}\right), & \;\mathrm{for}\;\alpha=\beta\\
\\
\frac{N_{\beta}}{M_{\alpha}}J_{\alpha\beta}\mathscr{A}_{\beta}'\left(\mu_{\beta}\right), & \;\mathrm{for}\;\alpha\neq\beta
\end{cases}\label{eq:reduced-Jacobian-matrix}
\end{equation}
}
\par\end{center}{\small \par}
\end{spacing}

\noindent and the corresponding eigenvectors $\boldsymbol{v}_{\alpha}^{\mathcal{R}}$
are:

\begin{spacing}{0.8}
\begin{center}
{\footnotesize{}
\begin{equation}
\boldsymbol{v}_{\alpha}^{\mathcal{R}}=\left[\begin{array}{c}
\left[\widehat{\boldsymbol{v}}_{\alpha}^{\mathcal{R}}\right]_{0}\\
\vdots\\
\left[\widehat{\boldsymbol{v}}_{\alpha}^{\mathcal{R}}\right]_{0}\\
\vdots\\
\quad\left[\widehat{\boldsymbol{v}}_{\alpha}^{\mathcal{R}}\right]_{\mathfrak{P}-1}\\
\vdots\\
\quad\left[\widehat{\boldsymbol{v}}_{\alpha}^{\mathcal{R}}\right]_{\mathfrak{P}-1}
\end{array}\right].\label{eq:Jacobian-matrix-eigenvectors-1}
\end{equation}
}
\par\end{center}{\footnotesize \par}
\end{spacing}

\noindent In \eqref{eq:Jacobian-matrix-eigenvectors-1} the entry
$\left[\widehat{\boldsymbol{v}}_{\alpha}^{\mathcal{R}}\right]_{\beta}$
(namely the $\beta$th entry of the vector $\widehat{\boldsymbol{v}}_{\alpha}^{\mathcal{R}}$)
is repeated $N_{\beta}$ times, and $\widehat{\boldsymbol{v}}_{\alpha}^{\mathcal{R}}=\left[\left[\widehat{\boldsymbol{v}}_{\alpha}^{\mathcal{R}}\right]_{0},\left[\widehat{\boldsymbol{v}}_{\alpha}^{\mathcal{R}}\right]_{1},\cdots,\left[\widehat{\boldsymbol{v}}_{\alpha}^{\mathcal{R}}\right]_{\mathfrak{P}-1}\right]^{T}$
is the eigenvector of $\mathcal{J}^{\mathcal{R}}$ corresponding to
the eigenvalue $\lambda_{\alpha}^{\mathcal{R}}$. Indeed, by developing
the product $\mathcal{J}\boldsymbol{v}_{\alpha}^{\mathcal{R}}$ the
equation $\mathcal{J}\boldsymbol{v}_{\alpha}^{\mathcal{R}}=\lambda_{\alpha}^{\mathcal{R}}\boldsymbol{v}_{\alpha}^{\mathcal{R}}$
can be rewritten as follows:

\begin{spacing}{0.8}
\begin{center}
{\small{}
\begin{equation}
\begin{cases}
\begin{array}{c}
{\displaystyle \sum_{p=0}^{N_{0}-1}}\mathcal{J}_{kp}\left[\boldsymbol{v}_{\alpha}^{\mathcal{R}}\right]_{p}+{\displaystyle \sum_{p=N_{0}}^{N_{0}+N_{1}-1}}\mathcal{J}_{kp}\left[\boldsymbol{v}_{\alpha}^{\mathcal{R}}\right]_{p}+...=\lambda_{\alpha}^{\mathcal{R}}\left[\boldsymbol{v}_{\alpha}^{\mathcal{R}}\right]_{k},\end{array} & k=0,...,N_{0}-1\\
\\
\qquad\qquad\qquad\qquad\vdots\\
\\
{\displaystyle \sum_{p=0}^{N_{0}-1}}\mathcal{J}_{kp}\left[\boldsymbol{v}_{\alpha}^{\mathcal{R}}\right]_{p}+{\displaystyle \sum_{p=N_{0}}^{N_{0}+N_{1}-1}}\mathcal{J}_{kp}\left[\boldsymbol{v}_{\alpha}^{\mathcal{R}}\right]_{p}+...=\lambda_{\alpha}^{\mathcal{R}}\left[\boldsymbol{v}_{\alpha}^{\mathcal{R}}\right]_{k}, & k=N_{0}+\ldots+N_{\mathfrak{P}-2},...,N-1
\end{cases}\label{eq:eigenvectors-system-0}
\end{equation}
}
\par\end{center}{\small \par}
\end{spacing}

\noindent Therefore, if we define:

\begin{spacing}{0.8}
\begin{center}
{\small{}
\begin{equation}
\mathfrak{S}_{\alpha\beta}\overset{\mathrm{def}}{=}{\displaystyle \sum_{k=n_{\beta-1}}^{n_{\beta}-1}}\left[\boldsymbol{v}_{\alpha}^{\mathcal{R}}\right]_{k}\label{eq:partial-sum-of-the-eigenvector-entries}
\end{equation}
}
\par\end{center}{\small \par}
\end{spacing}

\noindent for $\beta=0,\ldots,\mathfrak{P}-1$, and if we replace
the terms $\mathcal{J}_{kp}$ with the corresponding values given
by Eq.~\eqref{eq:Jacobian-matrix-multi-population}, the system \eqref{eq:eigenvectors-system-0}
can be rewritten as follows:

\begin{spacing}{0.8}
\begin{center}
{\small{}
\begin{equation}
\begin{array}{ccc}
{\displaystyle \sum_{\beta=0}^{\mathfrak{P}-1}}\frac{J_{\gamma\beta}}{M_{\gamma}}\mathscr{A}_{\beta}'\left(\mu_{\beta}\right)\mathfrak{S}_{\alpha\beta}=\left(\frac{1}{\tau_{\gamma}}+\frac{J_{\gamma\gamma}}{M_{\gamma}}\mathscr{A}_{\gamma}'\left(\mu_{\gamma}\right)+\lambda_{\alpha}^{\mathcal{R}}\right)\left[\boldsymbol{v}_{\alpha}^{\mathcal{R}}\right]_{k_{\gamma}}, &  & k_{\gamma}=n_{\gamma-1},...,n_{\gamma}-1\end{array}\label{eq:eigenvectors-system-1}
\end{equation}
}
\par\end{center}{\small \par}
\end{spacing}

\noindent for $\gamma=0,\ldots,\mathfrak{P}-1$. Now, Eq.~\eqref{eq:eigenvectors-system-1}
implies that $\left[\boldsymbol{v}_{\alpha}^{\mathcal{R}}\right]_{k_{\gamma}}$
does not depend on the index $k_{\gamma}$ for $\gamma$ fixed, therefore
from Eq.~\eqref{eq:partial-sum-of-the-eigenvector-entries} we get:

\begin{spacing}{0.8}
\begin{center}
{\small{}
\[
\mathfrak{S}_{\alpha\beta}=N_{\beta}\left[\boldsymbol{v}_{\alpha}^{\mathcal{R}}\right]_{k_{\beta}}.
\]
}
\par\end{center}{\small \par}
\end{spacing}

\noindent Therefore Eq.~\eqref{eq:eigenvectors-system-1} can be
rewritten as follows:

\begin{spacing}{0.8}
\begin{center}
{\small{}
\begin{equation}
{\displaystyle \sum_{\beta=0}^{\mathfrak{P}-1}}\frac{J_{\gamma\beta}}{M_{\gamma}}\mathscr{A}_{\beta}'\left(\mu_{\beta}\right)N_{\beta}\left[\boldsymbol{v}_{\alpha}^{\mathcal{R}}\right]_{k_{\beta}}=\left(\frac{1}{\tau_{\gamma}}+\frac{J_{\gamma\gamma}}{M_{\gamma}}\mathscr{A}_{\gamma}'\left(\mu_{\gamma}\right)+\lambda_{\alpha}^{\mathcal{R}}\right)\left[\boldsymbol{v}_{\alpha}^{\mathcal{R}}\right]_{k_{\gamma}}\label{eq:eigenvectors-system-2}
\end{equation}
}
\par\end{center}{\small \par}
\end{spacing}

\noindent This is a homogeneous system in the unknowns $\left[\boldsymbol{v}_{\alpha}^{\mathcal{R}}\right]_{k_{\beta}}$,
with coefficient matrix $\mathcal{J}^{\mathcal{R}}-\lambda_{\alpha}^{\mathcal{R}}\mathrm{Id}_{\mathfrak{P}}$,
where $\mathcal{J}^{\mathcal{R}}$ is given by Eq.~\eqref{eq:reduced-Jacobian-matrix}.
Therefore the system has a non-trivial solution if and only if $\det\left(\mathcal{J}^{\mathcal{R}}-\lambda_{\alpha}^{\mathcal{R}}\mathrm{Id}_{\mathfrak{P}}\right)=0$,
which defines the characteristic polynomial of the reduced Jacobian
matrix. The eigenvalues $\lambda_{\alpha}^{\mathcal{R}}$ are the
roots of this polynomial, with corresponding eigenvectors $\widehat{\boldsymbol{v}}_{\alpha}^{\mathcal{R}}\overset{\mathrm{def}}{=}\left[\left[\boldsymbol{v}_{\alpha}^{\mathcal{R}}\right]_{k_{0}},\left[\boldsymbol{v}_{\alpha}^{\mathcal{R}}\right]_{k_{1}},\cdots,\left[\boldsymbol{v}_{\alpha}^{\mathcal{R}}\right]_{k_{\mathfrak{P}-1}}\right]^{T}$. 

To conclude, we remind that the rank of the matrix $\mathcal{J}^{\mathcal{R}}-\lambda_{\alpha}^{\mathcal{R}}\mathrm{Id}_{\mathfrak{P}}$
determines the number of free components $\mathfrak{f}$ of $\boldsymbol{v}_{\alpha}^{\mathcal{R}}$,
through the relation $\mathfrak{f}=\mathfrak{P}-\mathrm{rank}\left(\mathcal{J}^{\mathcal{R}}-\lambda_{\alpha}^{\mathcal{R}}\mathrm{Id}_{\mathfrak{P}}\right)$
(rank-nullity theorem), with $0<\mathrm{rank}\left(\mathcal{J}^{\mathcal{R}}-\lambda_{\alpha}^{\mathcal{R}}\mathrm{Id}_{\mathfrak{P}}\right)<\mathfrak{P}$.

\subsection*{Example for $\mathfrak{P}=2$}

For simplicity, in the case of one excitatory and one inhibitory population,
we call the eigenvalues $\lambda_{0,1}^{\mathcal{P}}$ as $\lambda_{E,I}$,
and the corresponding eigenvectors $\boldsymbol{v}_{0,1}^{\mathcal{P}}$
as $\boldsymbol{v}_{E,I}$. Then, according to Eq.~\eqref{eq:Jacobian-matrix-degenerate-eigenvalues},
we get:

\begin{spacing}{0.8}
\begin{center}
{\small{}
\begin{equation}
\lambda_{E}=-\left[\frac{1}{\tau_{E}}+\frac{J_{EE}}{M_{E}}\mathscr{A}_{E}'\left(\mu_{E}\right)\right],\quad\lambda_{I}=-\left[\frac{1}{\tau_{I}}+\frac{J_{II}}{M_{I}}\mathscr{A}_{I}'\left(\mu_{I}\right)\right],\label{eq:two-populations-eigenvalues-0}
\end{equation}
}
\par\end{center}{\small \par}
\end{spacing}

\noindent with multiplicity $N_{E}-1$ and $N_{I}-1$ respectively,
while from Eq.~\eqref{eq:Jacobian-matrix-eigenvectors-0} we obtain:

\begin{spacing}{0.8}
\begin{center}
{\small{}
\begin{align}
 & \boldsymbol{v}_{E;0}=\left[\begin{array}{c}
1-N_{E}\\
1\\
\vdots\\
1\\
1\\
0\\
0\\
\vdots\\
0\\
0
\end{array}\right],\;\boldsymbol{v}_{E;1}=\left[\begin{array}{c}
1\\
1-N_{E}\\
\vdots\\
1\\
1\\
0\\
0\\
\vdots\\
0\\
0
\end{array}\right],\ldots,\boldsymbol{v}_{E;N_{E}-2}=\left[\begin{array}{c}
1\\
1\\
\vdots\\
1-N_{E}\\
1\\
0\\
0\\
\vdots\\
0\\
0
\end{array}\right],\nonumber \\
\label{eq:two-populations-eigenvectors-0}\\
 & \boldsymbol{v}_{I;0}=\left[\begin{array}{c}
0\\
0\\
\vdots\\
0\\
0\\
1-N_{I}\\
1\\
\vdots\\
1\\
1
\end{array}\right],\;\;\;\boldsymbol{v}_{I;1}=\left[\begin{array}{c}
0\\
0\\
\vdots\\
0\\
0\\
1\\
1-N_{I}\\
\vdots\\
1\\
1
\end{array}\right],\ldots,\;\;\;\boldsymbol{v}_{I;N_{I}-2}=\left[\begin{array}{c}
0\\
0\\
\vdots\\
0\\
0\\
1\\
1\\
\vdots\\
1-N_{I}\\
1
\end{array}\right].\nonumber 
\end{align}
}
\par\end{center}{\small \par}
\end{spacing}

\noindent The remaining eigenvalues of $\mathcal{J}$ are those of
the following reduced Jacobian matrix:

\begin{spacing}{0.8}
\begin{center}
{\small{}
\[
\mathcal{J}^{\mathcal{R}}=\left[\begin{array}{cc}
-\frac{1}{\tau_{E}}+\frac{N_{E}-1}{M_{E}}J_{EE}\mathscr{A}_{E}'\left(\mu_{E}\right) & \frac{N_{I}}{M_{E}}J_{EI}\mathscr{A}_{I}'\left(\mu_{I}\right)\\
\frac{N_{E}}{M_{I}}J_{IE}\mathscr{A}_{E}'\left(\mu_{E}\right) & -\frac{1}{\tau_{I}}+\frac{N_{I}-1}{M_{I}}J_{II}\mathscr{A}_{I}'\left(\mu_{I}\right)
\end{array}\right],
\]
}
\par\end{center}{\small \par}
\end{spacing}

\noindent namely:

\begin{spacing}{0.8}
\begin{center}
{\small{}
\begin{equation}
\lambda_{0,1}^{\mathcal{R}}=\frac{\mathcal{Y}+\mathcal{Z}\pm\sqrt{\left(\mathcal{Y}-\mathcal{Z}\right)^{2}+4\mathcal{X}}}{2},\label{q:two-populations-eigenvalues-1}
\end{equation}
}
\par\end{center}{\small \par}
\end{spacing}

\noindent where:

\begin{spacing}{0.8}
\begin{center}
{\small{}
\begin{equation}
\mathcal{X}=\frac{N_{E}N_{I}}{M_{E}M_{I}}J_{EI}J_{IE}\mathscr{A}_{E}'\left(\mu_{E}\right)\mathscr{A}_{I}'\left(\mu_{I}\right),\quad\mathcal{Y}=-\frac{1}{\tau_{E}}+\frac{N_{E}-1}{M_{E}}J_{EE}\mathscr{A}_{E}'\left(\mu_{E}\right),\quad\mathcal{Z}=-\frac{1}{\tau_{I}}+\frac{N_{I}-1}{M_{I}}J_{II}\mathscr{A}_{I}'\left(\mu_{I}\right).\label{eq:parameters-for-the-eigenvalues}
\end{equation}
}
\par\end{center}{\small \par}
\end{spacing}

\noindent We observe that $\mathrm{rank}\left(\mathcal{J}^{\mathcal{R}}-\lambda_{\alpha}^{\mathcal{R}}\mathrm{Id}_{\mathfrak{P}}\right)=1$,
therefore the eigenvectors of $\mathcal{J}^{\mathcal{R}}$ corresponding
to the eigenvalues $\lambda_{0,1}^{\mathcal{R}}$ are:

\begin{spacing}{0.8}
\begin{center}
{\footnotesize{}
\begin{equation}
\widehat{\boldsymbol{v}}_{0}^{\mathcal{R}}=x\left[\begin{array}{c}
1\\
K_{0}
\end{array}\right],\qquad\widehat{\boldsymbol{v}}_{1}^{\mathcal{R}}=y\left[\begin{array}{c}
1\\
K_{1}
\end{array}\right],\label{eq:reduced-Jacobian-matrix-eigenvectors}
\end{equation}
}
\par\end{center}{\footnotesize \par}
\end{spacing}

\noindent where:

\begin{spacing}{0.8}
\begin{center}
{\small{}
\begin{equation}
K_{\alpha}\overset{\mathrm{def}}{=}\frac{M_{E}\left(\lambda_{\alpha}^{\mathcal{R}}+\frac{1}{\tau_{E}}\right)-\left(N_{E}-1\right)J_{EE}\mathscr{A}_{E}'\left(\mu_{E}\right)}{N_{I}J_{EI}\mathscr{A}_{I}'\left(\mu_{I}\right)}=\frac{N_{E}J_{IE}\mathscr{A}_{E}'\left(\mu_{E}\right)}{M_{I}\left(\lambda_{\alpha}^{\mathcal{R}}+\frac{1}{\tau_{I}}\right)-\left(N_{I}-1\right)J_{II}\mathscr{A}_{I}'\left(\mu_{I}\right)}\label{eq:definition-of-the-terms-K}
\end{equation}
}
\par\end{center}{\small \par}
\end{spacing}

\noindent (the last equality is a consequence of $\det\left(\mathcal{J}^{\mathcal{R}}-\lambda_{\alpha}^{\mathcal{R}}\mathrm{Id}_{\mathfrak{P}}\right)=0$).
Moreover, $x$, $y$ are two free parameters that represent the free
components of $\widehat{\boldsymbol{v}}_{0,1}^{\mathcal{R}}$ (one
for each eigenvector, according to the relation $\mathfrak{f}=\mathfrak{P}-\mathrm{rank}\left(\mathcal{J}^{\mathcal{R}}-\lambda_{\alpha}^{\mathcal{R}}\mathrm{Id}_{\mathfrak{P}}\right)=1$).
Therefore the corresponding eigenvectors of $\mathcal{J}$ are:

\begin{spacing}{0.8}
\begin{center}
{\footnotesize{}
\begin{equation}
\boldsymbol{v}_{0}^{\mathcal{R}}=x\left[\begin{array}{c}
1\\
1\\
\vdots\\
1\\
K_{0}\\
K_{0}\\
\vdots\\
K_{0}
\end{array}\right],\qquad\boldsymbol{v}_{1}^{\mathcal{R}}=y\left[\begin{array}{c}
1\\
1\\
\vdots\\
1\\
K_{1}\\
K_{1}\\
\vdots\\
K_{1}
\end{array}\right],\label{eq:two-populations-eigenvectors-1}
\end{equation}
}
\par\end{center}{\footnotesize \par}
\end{spacing}

\noindent where the first $N_{E}$ entries of $\boldsymbol{v}_{0,1}^{\mathcal{R}}$
in Eq.~\eqref{eq:two-populations-eigenvectors-1} are equal to $1$
and thus $K_{0,1}$ are the remaining $N_{I}$ entries.

\section{Fundamental Matrix \label{sec:Fundamental-Matrix}}

When $\mathcal{J}$ is diagonalizable, its powers can be written as
$\mathcal{J}^{n}=PD^{n}P^{-1}$, where:

\begin{spacing}{0.8}
\begin{center}
{\small{}
\begin{align*}
D= & \mathrm{diag}\left(\lambda_{0}^{\mathcal{P}},\ldots,\lambda_{\mathfrak{P}-1}^{\mathcal{P}},\lambda_{0}^{\mathcal{R}},\ldots,\lambda_{\mathfrak{P}-1}^{\mathcal{R}}\right)\\
\\
D^{n}= & \mathrm{diag}\left(\left(\lambda_{0}^{\mathcal{P}}\right)^{n},\ldots,\left(\lambda_{\mathfrak{P}-1}^{\mathcal{P}}\right)^{n},\left(\lambda_{0}^{\mathcal{R}}\right)^{n},\ldots,\left(\lambda_{\mathfrak{P}-1}^{\mathcal{R}}\right)^{n}\right),
\end{align*}
}
\par\end{center}{\small \par}
\end{spacing}

\noindent while $P$ is an $N\times N$ matrix whose columns are composed
of the eigenvectors of $\mathcal{J}$ calculated in Sec.~\eqref{sec:Eigenvalues-and-Eigenvectors}.
If we introduce the matrices:

\begin{spacing}{0.8}
\begin{center}
{\small{}
\begin{equation}
\varPsi_{\alpha\beta;n}=\begin{cases}
\Xi_{\alpha;n}\mathrm{Id}{}_{N_{\alpha}}+\Gamma_{\alpha;n}\left(\mathbb{I}_{N_{\alpha}}-\mathrm{Id}_{N_{\alpha}}\right), & \;\mathrm{for}\;\alpha=\beta\\
\\
\Omega_{\alpha\beta;n}\mathbb{I}_{N_{\alpha},N_{\beta}}, & \;\mathrm{for}\;\alpha\neq\beta
\end{cases}\label{eq:Jacobian-matrix-powers-0}
\end{equation}
}
\par\end{center}{\small \par}
\end{spacing}

\noindent such that $\mathcal{J}^{n}=\left[\varPsi_{\alpha\beta;n}\right]_{\forall\left(\alpha,\beta\right)}$,
and we replace Eq.~\eqref{eq:Jacobian-matrix-powers-0} and the matrix
$P$ (expressed in terms of the eigenvectors of $\mathcal{J}$) into
the formula $\mathcal{J}^{n}P=PD^{n}$, after some algebra we get
the following set of equations:

\begin{spacing}{0.8}
\begin{center}
{\small{}
\begin{equation}
\begin{cases}
N_{\alpha}\left[\widehat{\boldsymbol{v}}_{\gamma}^{\mathcal{R}}\right]_{\alpha}\Gamma_{\alpha;n}+{\displaystyle \sum_{\substack{\beta=0\\
\beta\neq\alpha
}
}^{\mathfrak{P}-1}}N_{\beta}\left[\widehat{\boldsymbol{v}}_{\gamma}^{\mathcal{R}}\right]_{\beta}\Omega_{\alpha\beta;n}=\left[\widehat{\boldsymbol{v}}_{\gamma}^{\mathcal{R}}\right]_{\alpha}\left(\left(\lambda_{\gamma}^{\mathcal{R}}\right)^{n}-\left(\lambda_{\alpha}^{\mathcal{P}}\right)^{n}\right)\\
\\
\Xi_{\alpha;n}-\Gamma_{\alpha;n}=\left(\lambda_{\alpha}^{\mathcal{P}}\right)^{n}
\end{cases}\label{eq:Jacobian-matrix-power-system}
\end{equation}
}
\par\end{center}{\small \par}
\end{spacing}

\noindent for $\gamma=0,1,\ldots,\mathfrak{P}-1$. Now if we define:

\begin{spacing}{0.8}
\begin{center}
{\small{}
\begin{equation}
\Omega_{\alpha\alpha;n}\overset{\mathrm{def}}{=}\Gamma_{\alpha;n}+\frac{1}{N_{\alpha}}\left(\lambda_{\alpha}^{\mathcal{P}}\right)^{n},\label{eq:definition-of-the-omega-terms}
\end{equation}
}
\par\end{center}{\small \par}
\end{spacing}

\noindent the first equation of the system \eqref{eq:Jacobian-matrix-power-system}
becomes:

\begin{spacing}{0.8}
\begin{center}
{\small{}
\[
{\displaystyle \sum_{\beta=0}^{\mathfrak{P}-1}}N_{\beta}\left[\widehat{\boldsymbol{v}}_{\gamma}^{\mathcal{R}}\right]_{\beta}\Omega_{\alpha\beta;n}=\left[\widehat{\boldsymbol{v}}_{\gamma}^{\mathcal{R}}\right]_{\alpha}\left(\lambda_{\gamma}^{\mathcal{R}}\right)^{n},\quad\gamma=0,1,\ldots,\mathfrak{P}-1.
\]
}
\par\end{center}{\small \par}
\end{spacing}

\noindent In matrix form the system reads $\mathfrak{A}\boldsymbol{\mathfrak{x}}_{\alpha;n}=\boldsymbol{\mathfrak{b}}_{\alpha;n}$,
where:

\begin{spacing}{0.8}
\begin{center}
{\small{}
\begin{align}
 & \mathfrak{A}=\left(P^{\mathcal{R}}\right)^{T}\mathrm{diag}\left(N_{0},N_{1},\cdots,N_{\mathfrak{P}-1}\right)\nonumber \\
\label{eq:system-of-linear-equations-for-the-terms-omega}\\
 & \boldsymbol{\mathfrak{x}}_{\alpha;n}=\left[\begin{array}{c}
\Omega_{\alpha0;n}\\
\Omega_{\alpha1;n}\\
\vdots\\
\Omega_{\alpha,\mathfrak{P}-1;n}
\end{array}\right],\quad\boldsymbol{\mathfrak{b}}_{\alpha;n}=\left[\begin{array}{c}
\left[\widehat{\boldsymbol{v}}_{0}^{\mathcal{R}}\right]_{\alpha}\left(\lambda_{0}^{\mathcal{R}}\right)^{n}\\
\left[\widehat{\boldsymbol{v}}_{1}^{\mathcal{R}}\right]_{\alpha}\left(\lambda_{1}^{\mathcal{R}}\right)^{n}\\
\vdots\\
\left[\widehat{\boldsymbol{v}}_{\mathfrak{P}-1}^{\mathcal{R}}\right]_{\alpha}\left(\lambda_{\mathfrak{P}-1}^{\mathcal{R}}\right)^{n}
\end{array}\right],\quad\alpha=0,1,\ldots,\mathfrak{P}-1.\nonumber 
\end{align}
}
\par\end{center}{\small \par}
\end{spacing}

\noindent In Eq.~\eqref{eq:system-of-linear-equations-for-the-terms-omega},
$P^{\mathcal{R}}$ is a $\mathfrak{P}\times\mathfrak{P}$ matrix whose
columns are composed of the eigenvectors of $\mathcal{J}^{\mathcal{R}}$,
namely:

\begin{spacing}{0.8}
\begin{center}
{\small{}
\[
P^{\mathcal{R}}=\left[\begin{array}{cccc}
\left[\widehat{\boldsymbol{v}}_{0}^{\mathcal{R}}\right]_{0} & \left[\widehat{\boldsymbol{v}}_{1}^{\mathcal{R}}\right]_{0} & \cdots & \left[\widehat{\boldsymbol{v}}_{\mathfrak{P}-1}^{\mathcal{R}}\right]_{0}\\
\left[\widehat{\boldsymbol{v}}_{0}^{\mathcal{R}}\right]_{1} & \left[\widehat{\boldsymbol{v}}_{1}^{\mathcal{R}}\right]_{1} & \cdots & \left[\widehat{\boldsymbol{v}}_{\mathfrak{P}-1}^{\mathcal{R}}\right]_{1}\\
\vdots & \vdots & \ddots & \vdots\\
\quad\left[\widehat{\boldsymbol{v}}_{0}^{\mathcal{R}}\right]_{\mathfrak{P}-1} & \quad\left[\widehat{\boldsymbol{v}}_{1}^{\mathcal{R}}\right]_{\mathfrak{P}-1} & \cdots & \quad\left[\widehat{\boldsymbol{v}}_{\mathfrak{P}-1}^{\mathcal{R}}\right]_{\mathfrak{P}-1}
\end{array}\right].
\]
}
\par\end{center}{\small \par}
\end{spacing}

\noindent Therefore by inverting the system $\mathfrak{A}\boldsymbol{\mathfrak{x}}_{\alpha;n}=\boldsymbol{\mathfrak{b}}_{\alpha;n}$
we obtain:

\begin{spacing}{0.8}
\begin{center}
{\small{}
\[
\Omega_{\alpha\beta;n}={\displaystyle \sum_{\gamma=0}^{\mathfrak{P}-1}}\mathfrak{C}_{\alpha\beta}^{\left(\gamma\right)}\left(\lambda_{\gamma}^{\mathcal{R}}\right)^{n},
\]
}
\par\end{center}{\small \par}
\end{spacing}

\noindent where:

\begin{spacing}{0.8}
\begin{center}
{\small{}
\begin{equation}
\mathfrak{C}_{\alpha\beta}^{\left(\gamma\right)}=\left[\mathfrak{A}^{-1}\right]_{\beta\gamma}\left[\widehat{\boldsymbol{v}}_{\gamma}^{\mathcal{R}}\right]_{\alpha}=\frac{1}{N_{\beta}}\left[\left(P^{\mathcal{R}}\right)^{-T}\right]_{\beta\gamma}\left[\widehat{\boldsymbol{v}}_{\gamma}^{\mathcal{R}}\right]_{\alpha},\quad\left(P^{\mathcal{R}}\right)^{-T}\overset{\mathrm{def}}{=}\left[\left(P^{\mathcal{R}}\right)^{T}\right]^{-1}.\label{eq:coefficients-C-0}
\end{equation}
}
\par\end{center}{\small \par}
\end{spacing}

\noindent We observe that the matrix $\mathfrak{C}^{\left(\gamma\right)}\overset{\mathrm{def}}{=}\left[\mathfrak{C}_{\alpha\beta}^{\left(\gamma\right)}\right]_{\forall\left(\alpha,\beta\right)}$
can be written as an outer product of vectors, namely:

\begin{spacing}{0.8}
\begin{center}
{\small{}
\begin{equation}
\mathfrak{C}^{\left(\gamma\right)}=\left[\begin{array}{c}
\left[\widehat{\boldsymbol{v}}_{\gamma}^{\mathcal{R}}\right]_{0}\\
\left[\widehat{\boldsymbol{v}}_{\gamma}^{\mathcal{R}}\right]_{1}\\
\vdots\\
\quad\left[\widehat{\boldsymbol{v}}_{\gamma}^{\mathcal{R}}\right]_{\mathfrak{P}-1}
\end{array}\right]\left[\left[\mathfrak{A}^{-1}\right]_{0\gamma},\left[\mathfrak{A}^{-1}\right]_{1\gamma},\cdots,\left[\mathfrak{A}^{-1}\right]_{\mathfrak{P}-1,\gamma}\right],\label{eq:coefficients-C-1}
\end{equation}
}
\par\end{center}{\small \par}
\end{spacing}

\noindent therefore it has rank one. This result will prove very important
in demonstrating the formation of critical slowing down near the saddle-node
bifurcations (see SubSec.~\eqref{sub:Saddle-Node-Bifurcations}).

Moreover, from the second equation of \eqref{eq:Jacobian-matrix-power-system}
and from the definition \eqref{eq:definition-of-the-omega-terms},
we get:

\begin{spacing}{0.8}
\begin{center}
{\small{}
\[
\Xi_{\alpha;n}=\left(\lambda_{\alpha}^{\mathcal{P}}\right)^{n}+\Gamma_{\alpha;n}=\left(1-\frac{1}{N_{\alpha}}\right)\left(\lambda_{\alpha}^{\mathcal{P}}\right)^{n}+\Omega_{\alpha\alpha;n},
\]
}
\par\end{center}{\small \par}
\end{spacing}

\noindent Therefore we can rewrite Eq.~\eqref{eq:Jacobian-matrix-powers-0}
as follows:

\begin{spacing}{0.8}
\begin{center}
{\small{}
\begin{equation}
\varPsi_{\alpha\beta;n}=\begin{cases}
\left({\displaystyle \sum_{\gamma=0}^{\mathfrak{P}-1}}\mathfrak{C}_{\alpha\alpha}^{\left(\gamma\right)}\left(\lambda_{\gamma}^{\mathcal{R}}\right)^{n}\right)\mathbb{I}_{N_{\alpha}}+\left(-\frac{1}{N_{\alpha}}\mathbb{I}_{N_{\alpha}}+\mathrm{Id}{}_{N_{\alpha}}\right)\left(\lambda_{\alpha}^{\mathcal{P}}\right)^{n}, & \;\mathrm{for}\;\alpha=\beta\\
\\
\left({\displaystyle \sum_{\gamma=0}^{\mathfrak{P}-1}}\mathfrak{C}_{\alpha\beta}^{\left(\gamma\right)}\left(\lambda_{\gamma}^{\mathcal{R}}\right)^{n}\right)\mathbb{I}_{N_{\alpha},N_{\beta}}, & \;\mathrm{for}\;\alpha\neq\beta.
\end{cases}\label{eq:Jacobian-matrix-powers-1}
\end{equation}
}
\par\end{center}{\small \par}
\end{spacing}

To conclude, from the Taylor expansion $e^{\mathcal{J}t}={\displaystyle \sum_{n=0}^{\infty}\frac{t^{n}}{n!}\mathcal{J}^{n}}$
and Eq.~\eqref{eq:Jacobian-matrix-powers-1} we get $\Phi\left(t\right)=\left[\varPhi_{\alpha\beta}\left(t\right)\right]_{\forall\left(\alpha,\beta\right)}$,
where:

\begin{spacing}{0.8}
\begin{center}
{\small{}
\begin{equation}
\varPhi_{\alpha\beta}\left(t\right)=\begin{cases}
\left({\displaystyle \sum_{\gamma=0}^{\mathfrak{P}-1}}\mathfrak{C}_{\alpha\alpha}^{\left(\gamma\right)}e^{\lambda_{\gamma}^{\mathcal{R}}t}\right)\mathbb{I}_{N_{\alpha}}+\left(-\frac{1}{N_{\alpha}}\mathbb{I}_{N_{\alpha}}+\mathrm{Id}_{N_{\alpha}}\right)e^{\lambda_{\alpha}^{\mathcal{P}}t}, & \;\mathrm{for}\;\alpha=\beta\\
\\
\left({\displaystyle \sum_{\gamma=0}^{\mathfrak{P}-1}}\mathfrak{C}_{\alpha\beta}^{\left(\gamma\right)}e^{\lambda_{\gamma}^{\mathcal{R}}t}\right)\mathbb{I}_{N_{\alpha},N_{\beta}}, & \;\mathrm{for}\;\alpha\neq\beta.
\end{cases}\label{eq:fundamental-matrix}
\end{equation}
}
\par\end{center}{\small \par}
\end{spacing}

\noindent For the sake of clarity, we implemented this method in the
supplemental Python script ``Fundamental\_Matrix.py'' for an arbitrary
number of populations. Furthermore, below we show the explicit calculation
of the fundamental matrix in the case $\mathfrak{P}=2$.

\subsection*{Example for $\mathfrak{P}=2$}

In the case of two neural populations, from Eqs.~\eqref{eq:reduced-Jacobian-matrix-eigenvectors}
we get $P^{\mathcal{R}}=\left[\begin{array}{cc}
1 & 1\\
K_{0} & K_{1}
\end{array}\right]$ and therefore $\left(P^{\mathcal{R}}\right)^{-T}=\frac{1}{K_{1}-K_{0}}\left[\begin{array}{cc}
K_{1} & -K_{0}\\
-1 & 1
\end{array}\right]$. Thus, according to Eqs.~\eqref{eq:coefficients-C-0} + \eqref{eq:fundamental-matrix},
the blocks of the fundamental matrix $\Phi\left(t\right)=\left[\begin{array}{cc}
\varPhi_{EE}\left(t\right) & \varPhi_{EI}\left(t\right)\\
\varPhi_{IE}\left(t\right) & \varPhi_{II}\left(t\right)
\end{array}\right]$ are given by the following formulas:

\begin{spacing}{0.8}
\begin{center}
{\small{}
\begin{align}
 & \varPhi_{EE}\left(t\right)=\frac{K_{1}e^{\lambda_{0}^{\mathcal{R}}t}-K_{0}e^{\lambda_{1}^{\mathcal{R}}t}}{N_{E}\left(K_{1}-K_{0}\right)}\mathbb{I}_{N_{E}}+\left(-\frac{1}{N_{E}}\mathbb{I}_{N_{E}}+\mathrm{Id}_{N_{E}}\right)e^{\lambda_{E}t}, &  & \varPhi_{EI}\left(t\right)=\frac{e^{\lambda_{1}^{\mathcal{R}}t}-e^{\lambda_{0}^{\mathcal{R}}t}}{N_{I}\left(K_{1}-K_{0}\right)}\mathbb{I}_{N_{E},N_{I}},\nonumber \\
\label{eq:fundamental-matrix-two-populations-case}\\
 & \varPhi_{II}\left(t\right)=\frac{K_{1}e^{\lambda_{1}^{\mathcal{R}}t}-K_{0}e^{\lambda_{0}^{\mathcal{R}}t}}{N_{I}\left(K_{1}-K_{0}\right)}\mathbb{I}_{N_{I}}+\left(-\frac{1}{N_{I}}\mathbb{I}_{N_{I}}+\mathrm{Id}_{N_{I}}\right)e^{\lambda_{I}t}, &  & \varPhi_{IE}\left(t\right)=\frac{K_{0}K_{1}\left(e^{\lambda_{0}^{\mathcal{R}}t}-e^{\lambda_{1}^{\mathcal{R}}t}\right)}{N_{E}\left(K_{1}-K_{0}\right)}\mathbb{I}_{N_{I},N_{E}}.\nonumber 
\end{align}
}
\par\end{center}{\small \par}
\end{spacing}

\section{Constraints for the Covariance Matrix of the Noise \label{sec:Constraints-for-the-Covariance-Matrix-of-the-Noise}}

In order to be a genuine covariance matrix, $\Sigma^{\mathscr{B}}$
(see Eq.~(4) in the main text) must be positive-semidefinite. Being
symmetric, $\Sigma^{\mathscr{B}}$ is positive-semidefinite if and
only if its eigenvalues are non-negative. Since it has the same block
structure of the Jacobian matrix, we can easily calculate its eigenvalues
by adapting the results of Sec.~\eqref{sec:Eigenvalues-and-Eigenvectors}.
For example, in the case $\mathfrak{P}=2$ we get that the matrix
$\Sigma^{\mathscr{B}}$ has to satisfy the following set of inequalities
in order to be positive-semidefinite:

\begin{spacing}{0.8}
\begin{center}
{\small{}
\[
\begin{cases}
\left(\sigma_{E}^{\mathscr{B}}\right)^{2}\left[1+\left(N_{E}-1\right)C_{EE}^{\mathscr{B}}\right]+\left(\sigma_{I}^{\mathscr{B}}\right)^{2}\left[1+\left(N_{I}-1\right)C_{II}^{\mathscr{B}}\right]\geq0\\
\\
\left[1+\left(N_{E}-1\right)C_{EE}^{\mathscr{B}}\right]\left[1+\left(N_{I}-1\right)C_{II}^{\mathscr{B}}\right]\geq N_{E}N_{I}\left(C_{EI}^{\mathscr{B}}\right)^{2}.
\end{cases}
\]
}
\par\end{center}{\small \par}
\end{spacing}

\noindent The importance of this constraint is highlighted by the
following relation:

\begin{spacing}{0.8}
\begin{center}
{\small{}
\[
{\displaystyle \sum_{i,j=0}^{N-1}}x_{i}x_{j}\mathrm{Cov}\left(\frac{d\mathscr{B}_{i}\left(t\right)}{dt},\frac{d\mathscr{B}_{j}\left(t\right)}{dt}\right)=\mathrm{Var}\left({\displaystyle \sum_{i=0}^{N-1}}x_{i}\frac{d\mathscr{B}_{i}\left(t\right)}{dt}\right).
\]
}
\par\end{center}{\small \par}
\end{spacing}

\noindent Indeed, from this equality we see that if $\Sigma^{\mathscr{B}}$
were not positive-semidefinite, the linear combination ${\displaystyle \sum_{i=0}^{N-1}}x_{i}\frac{d\mathscr{B}_{i}\left(t\right)}{dt}$
would have negative variance for some coefficients $x_{i}$.

\section{Functional Connectivity \label{sec:Functional-Connectivity}}

\noindent Now we have all the ingredients for studying the phenomena
that affects the functional connectivity of the network. In particular,
in this section we prove that, depending on the parameters of the
network, the neural activity spans from strong anti-correlation (i.e.
$\mathrm{Corr}\left(V_{i}\left(t\right),V_{j}\left(t\right)\right)\rightarrow-1$)
to strongly positive correlation ($\mathrm{Corr}\left(V_{i}\left(t\right),V_{j}\left(t\right)\right)\rightarrow1$),
passing through arbitrarily weak correlation ($\mathrm{Corr}\left(V_{i}\left(t\right),V_{j}\left(t\right)\right)\rightarrow0$).
We consider the weak-correlation regime in SubSec.~\eqref{sub:Weak-Correlation-Regime-(Local-Chaos)},
while we study the strong-correlation regime in SubSec.~\eqref{sub:Strong-Correlation-Regime-(Local-Bifurcations)}
in relation to the local bifurcations of the network. For both the
regimes, we analyze the variance and the cross-correlation for an
arbitrary number of populations, extending the results of the case
$\mathfrak{P}=2$ that we developed in the main text (see Eqs.~(14)
+ (15)) without the need of explicit formulas for the functional connectivity.
Indeed, we show that it is possible to evaluate the qualitative behavior
of the covariance matrix of the membrane potentials $\Sigma^{V}$
even if the explicit expressions of the coefficients $\mathfrak{C}_{\alpha\beta}^{\left(\gamma\right)}$
in Eq.~\eqref{eq:fundamental-matrix} are not known.

\subsection{Weak-Correlation Regime (Local Chaos) \label{sub:Weak-Correlation-Regime-(Local-Chaos)}}

According to the mean-field theory developed by McKean, Tanaka and
Sznitman \cite{McKean1966,McKean1967,Tanaka1978,Tanaka1981,Tanaka1983,Sznitman1984,Sznitman1986,Sznitman1991},
the cross-correlation of the multi-population network in the thermodynamic
limit ($N_{\alpha}\rightarrow\infty$ $\forall\alpha$) tends to zero
for every pair of neurons if the Brownian motions are independent
(i.e. $C_{\alpha\beta}^{\mathscr{B}}=0\;\forall\alpha,\beta$). This
result was proved in \cite{Touboul2012} for an arbitrary number of
neural populations with infinite size.

However, having an infinite size is not the only way a network can
experience low levels of correlation. As explained in \cite{Fasoli2015},
for strongly depolarizing or strongly hyperpolarizing stimuli the
activation function saturates to $\nu^{\mathrm{max}}$ or $0$ respectively
(see Eq.~(2) of the main text), therefore $\mathscr{A}'\left(V\right)\rightarrow0$.
Given a network with an arbitrary number of populations of arbitrary
size, if the stimulus is strong enough for all the populations, the
Jacobian matrix of the network (see Eq.~\eqref{eq:Jacobian-matrix-multi-population})
becomes diagonal. In other words, the neurons become \textsl{effectively}
disconnected. For this reason the correlation structure of the membrane
potentials is determined only by the Brownian motions, namely $\mathrm{Corr}\left(V_{i}\left(t\right),V_{j}\left(t\right)\right)\rightarrow\left[\Sigma^{\mathscr{B}}\right]_{ij}$.
Therefore, the necessary requirement to generate weak correlation
between neurons is again the absence of noise correlations, namely
$C_{\alpha\beta}^{\mathscr{B}}=0\;\forall\alpha,\beta$. This mechanism
is different from that described in \cite{Touboul2012}, since it
occurs for $\left|I_{\alpha}\right|\rightarrow\infty$ rather than
$N_{\alpha}\rightarrow\infty$, and it allows the network to create
decorrelated activity even if its size is small.

\subsection{Strong-Correlation Regime (Local Bifurcations) \label{sub:Strong-Correlation-Regime-(Local-Bifurcations)}}

Interestingly, the network is able to show also high values of correlation,
either positive or negative. This phenomenon occurs at the (local)
bifurcation points of the network, and therefore for special combinations
of the network's parameters. As we explained in the main text, we
consider only the codimension one bifurcations of the network, namely
saddle-node, Andronov-Hopf and branching-point bifurcations. The relation
between these bifurcations and the functional connectivity of the
system is described in the next subsections.

\subsubsection{Saddle-Node Bifurcations \label{sub:Saddle-Node-Bifurcations}}

The multi-population neural network undergoes saddle-node bifurcations
when one of the eigenvalues of the reduced Jacobian matrix tends to
zero \cite{Fasoli2016}. Therefore, if $\lambda_{\gamma}^{\mathcal{R}}\rightarrow0^{-}$
for a given $\gamma$, and all the other eigenvalues have negative
real part, then for $t\gg\frac{1}{\left|\lambda_{\gamma}^{\mathcal{R}}\right|}$
Eq.~\eqref{eq:fundamental-matrix} gives:

\begin{spacing}{0.8}
\begin{center}
{\small{}
\[
\varPhi_{\alpha\beta}\left(t\right)\approx\mathfrak{C}_{\alpha\beta}^{\left(\gamma\right)}e^{-\left|\lambda_{\gamma}^{\mathcal{R}}\right|t}\mathbb{I}_{N_{\alpha},N_{\beta}}\quad\forall\alpha,\beta.
\]
}
\par\end{center}{\small \par}
\end{spacing}

\noindent According to Eq.~(6) of the main text, this implies:

\begin{spacing}{0.8}
\begin{center}
{\small{}
\begin{equation}
\mathrm{Cov}\left(V_{i}\left(t\right),V_{j}\left(t\right)\right)\approx\frac{1-e^{-2\left|\lambda_{\gamma}^{\mathcal{R}}\right|t}}{2\left|\lambda_{\gamma}^{\mathcal{R}}\right|}\sum_{\beta=0}^{\mathfrak{P}-1}N_{\beta}\left(\sigma_{\beta}^{\mathscr{B}}\right)^{2}\mathfrak{C}_{\mathfrak{p}\left(i\right),\beta}^{\left(\gamma\right)}\mathfrak{C}_{\mathfrak{p}\left(j\right),\beta}^{\left(\gamma\right)},\;\forall i,j\label{eq:saddle-node-covariance}
\end{equation}
}
\par\end{center}{\small \par}
\end{spacing}

\noindent where $\mathfrak{p}\left(k\right)$ represents the neural
population the $k$th neuron belongs to. Now, since $\mathrm{Var}\left(V_{i}\left(t\right)\right)=\mathrm{Cov}\left(V_{i}\left(t\right),V_{i}\left(t\right)\right)\approx\frac{1}{2\left|\lambda_{\gamma}^{\mathcal{R}}\right|}\sum_{\beta=0}^{\mathfrak{P}-1}N_{\beta}\left(\sigma_{\beta}^{\mathscr{B}}\mathfrak{C}_{\mathfrak{p}\left(i\right),\beta}^{\left(\gamma\right)}\right)^{2}$
for $t\gg\frac{1}{\left|\lambda_{\gamma}^{\mathcal{R}}\right|}$,
the amplitude of the fluctuations diverges for $\lambda_{\gamma}^{\mathcal{R}}\rightarrow0^{-}$.
Moreover, if $\mathfrak{p}\left(i\right)=\mathfrak{p}\left(j\right)$,
Eq.~\eqref{eq:saddle-node-covariance} gives $\mathrm{Cov}\left(V_{i}\left(t\right),V_{j}\left(t\right)\right)\approx\mathrm{Var}\left(V_{i}\left(t\right)\right)=\mathrm{Var}\left(V_{j}\left(t\right)\right)$,
therefore according to Pearson's formula $\mathrm{Corr}\left(V_{i}\left(t\right),V_{j}\left(t\right)\right)\approx1$.
On the other side, if $\mathfrak{p}\left(i\right)\neq\mathfrak{p}\left(j\right)$,
by replacing Eq.~\eqref{eq:saddle-node-covariance} within the condition
$\mathrm{Cov}^{2}\left(V_{i}\left(t\right),V_{j}\left(t\right)\right)=\mathrm{Var}\left(V_{i}\left(t\right)\right)\mathrm{Var}\left(V_{j}\left(t\right)\right)$,
after some algebra we get:

\begin{spacing}{0.8}
\begin{center}
{\small{}
\[
\sum_{\begin{array}{c}
\beta_{0},\beta_{1}=0\\
\beta_{0}<\beta_{1}
\end{array}}^{\mathfrak{P}-1}N_{\beta_{0}}N_{\beta_{1}}\left(\sigma_{\beta_{0}}^{\mathscr{B}}\sigma_{\beta_{1}}^{\mathscr{B}}\right)^{2}\left(\mathfrak{C}_{\mathfrak{p}\left(i\right),\beta_{0}}^{\left(\gamma\right)}\mathfrak{C}_{\mathfrak{p}\left(j\right),\beta_{1}}^{\left(\gamma\right)}-\mathfrak{C}_{\mathfrak{p}\left(i\right),\beta_{1}}^{\left(\gamma\right)}\mathfrak{C}_{\mathfrak{p}\left(j\right),\beta_{0}}^{\left(\gamma\right)}\right)^{2}=0.
\]
}
\par\end{center}{\small \par}
\end{spacing}

\noindent This equality is satisfied if and only if:

\begin{spacing}{0.8}
\begin{center}
{\small{}
\[
\mathfrak{C}_{\mathfrak{p}\left(i\right),\beta_{0}}^{\left(\gamma\right)}\mathfrak{C}_{\mathfrak{p}\left(j\right),\beta_{1}}^{\left(\gamma\right)}-\mathfrak{C}_{\mathfrak{p}\left(i\right),\beta_{1}}^{\left(\gamma\right)}\mathfrak{C}_{\mathfrak{p}\left(j\right),\beta_{0}}^{\left(\gamma\right)}=\det\left(\left[\begin{array}{cc}
\mathfrak{C}_{\mathfrak{p}\left(i\right),\beta_{0}}^{\left(\gamma\right)} & \mathfrak{C}_{\mathfrak{p}\left(i\right),\beta_{1}}^{\left(\gamma\right)}\\
\mathfrak{C}_{\mathfrak{p}\left(j\right),\beta_{0}}^{\left(\gamma\right)} & \mathfrak{C}_{\mathfrak{p}\left(j\right),\beta_{1}}^{\left(\gamma\right)}
\end{array}\right]\right)=0\quad\forall\beta_{0},\beta_{1},
\]
}
\par\end{center}{\small \par}
\end{spacing}

\noindent namely if and only if the matrix $\mathfrak{C}^{\left(\gamma\right)}$
has rank $1$, which is indeed the case (see Sec.~\eqref{sec:Fundamental-Matrix}).
This proves the emergence of strong correlations also between the
neurons of different populations when the network is close to a saddle-node
bifurcation.

Interestingly, critical slowing down occurs regardless of the strength
of the correlation between the Brownian motions.

\subsubsection{Andronov-Hopf Bifurcations \label{sub:Andronov-Hopf-Bifurcations}}

The multi-population neural network undergoes an Andronov-Hopf bifurcation
whenever the reduced Jacobian matrix has two complex-conjugate purely
imaginary eigenvalues \cite{Fasoli2016}. If the real part of a pair
of complex-conjugate eigenvalues $\lambda_{\gamma,\delta}^{\mathcal{R}}$
tends to zero (i.e. $\mathfrak{R}\left(\lambda_{\gamma}^{\mathcal{R}}\right)\rightarrow0^{-}$),
and if all the other eigenvalues have negative real part, then for
$t\gg\frac{1}{\left|\mathfrak{R}\left(\lambda_{\gamma}^{\mathcal{R}}\right)\right|}$
Eq.~\eqref{eq:fundamental-matrix} gives:

\begin{spacing}{0.8}
\begin{center}
{\small{}
\[
\varPhi_{\alpha\beta}\left(t\right)\approx\left(\mathfrak{C}_{\alpha\beta}^{\left(\gamma\right)}e^{\lambda_{\gamma}^{\mathcal{R}}t}+\overline{\mathfrak{C}}_{\alpha\beta}^{\left(\gamma\right)}e^{\overline{\lambda}_{\gamma}^{\mathcal{R}}t}\right)\mathbb{I}_{N_{\alpha},N_{\beta}}\quad\forall\alpha,\beta
\]
}
\par\end{center}{\small \par}
\end{spacing}

\noindent where $\overline{\lambda}_{\gamma}^{\mathcal{R}}=\lambda_{\delta}^{\mathcal{R}}$
and the overline represents the complex conjugate operator. According
to Eq.~(6) of the main text, after some algebra we get:

\begin{spacing}{0.8}
\begin{center}
{\small{}
\begin{equation}
\mathrm{Cov}\left(V_{i}\left(t\right),V_{j}\left(t\right)\right)\sim\frac{1-e^{-2\left|\mathfrak{R}\left(\lambda_{\gamma}^{\mathcal{R}}\right)\right|t}}{\left|\mathfrak{R}\left(\lambda_{\gamma}^{\mathcal{R}}\right)\right|}\sum_{\beta=0}^{\mathfrak{P}-1}N_{\beta}\left(\sigma_{\beta}^{\mathscr{B}}\right)^{2}\mathfrak{R}\left(\mathfrak{C}_{\mathfrak{p}\left(i\right),\beta}^{\left(\gamma\right)}\overline{\mathfrak{C}}_{\mathfrak{p}\left(j\right),\beta}^{\left(\gamma\right)}\right),\;\forall i,j\label{eq:Hopf-covariance}
\end{equation}
}
\par\end{center}{\small \par}
\end{spacing}

\noindent thus, similarly to the saddle-node bifurcations, for $t\gg\frac{1}{\left|\mathfrak{R}\left(\lambda_{\gamma}^{\mathcal{R}}\right)\right|}$
the amplitude of the fluctuations diverges for $\mathfrak{R}\left(\lambda_{\gamma}^{\mathcal{R}}\right)\rightarrow0^{-}$.
Moreover, if $\mathfrak{p}\left(i\right)=\mathfrak{p}\left(j\right)$,
Eq.~\eqref{eq:Hopf-covariance} gives $\mathrm{Cov}\left(V_{i}\left(t\right),V_{j}\left(t\right)\right)\approx\mathrm{Var}\left(V_{i}\left(t\right)\right)=\mathrm{Var}\left(V_{j}\left(t\right)\right)$,
therefore $\mathrm{Corr}\left(V_{i}\left(t\right),V_{j}\left(t\right)\right)\approx1$.
On the other side, if $\mathfrak{p}\left(i\right)\neq\mathfrak{p}\left(j\right)$,
from Eq.~\eqref{eq:Hopf-covariance} it is possible to prove that
the condition $\mathrm{Cov}^{2}\left(V_{i}\left(t\right),V_{j}\left(t\right)\right)=\mathrm{Var}\left(V_{i}\left(t\right)\right)\mathrm{Var}\left(V_{j}\left(t\right)\right)$
is satisfied if and only if $\mathfrak{I}\left(\mathfrak{C}_{\mathfrak{p}\left(i\right),\beta}^{\left(\gamma\right)}\right)=\mathfrak{I}\left(\mathfrak{C}_{\mathfrak{p}\left(j\right),\beta}^{\left(\gamma\right)}\right)=0\;\forall\beta$.
However, these imaginary parts cannot be equal to zero at an Andronov-Hopf
bifurcation, the latter being defined by a pair of purely imaginary
eigenvalues. Therefore, contrary to the saddle-node bifurcations,
the cross-correlation between neurons in different populations does
not tend to one at an Andronov-Hopf bifurcation. This is also confirmed
by the top-right panel of Fig.~(4) in the main text, in the special
case $\mathfrak{P}=2$.

\subsubsection{Branching-Point Bifurcations \label{sub:Branching-Point-Bifurcations}}

Branching-point bifurcations occur whenever $\lambda_{\gamma}^{\mathcal{P}}\rightarrow0^{-}$
for a given $\gamma$ \cite{Fasoli2016}. According to Eq.~\eqref{eq:Jacobian-matrix-degenerate-eigenvalues},
only sufficiently strong self-inhibitory synaptic weights $J_{\gamma\gamma}$
give rise to this kind of bifurcation. As we explained in \cite{Fasoli2016},
branching points do not occur in the mean-field approximation of the
network, therefore the study of the functional connectivity near these
bifurcations is one of the results of most interest of our article.

If $\lambda_{\gamma}^{\mathcal{P}}\rightarrow0^{-}$ and all the other
eigenvalues have negative real part, for $t\gg\frac{1}{\left|\lambda_{\gamma}^{\mathcal{P}}\right|}$
Eq.~\eqref{eq:fundamental-matrix} gives:

\begin{spacing}{0.8}
\begin{center}
{\small{}
\begin{equation}
\varPhi_{\gamma\beta}\left(t\right)\approx\begin{cases}
\left(-\frac{1}{N_{\gamma}}\mathbb{I}_{N_{\gamma}}+\mathrm{Id}_{N_{\gamma}}\right)e^{-\left|\lambda_{\gamma}^{\mathcal{P}}\right|t}, & \;\mathrm{for}\;\gamma=\beta\\
\\
\mathbb{O}_{N_{\gamma},N_{\beta}}, & \;\mathrm{for}\;\gamma\neq\beta
\end{cases}\label{eq:fundamental-matrix-branching-point}
\end{equation}
}
\par\end{center}{\small \par}
\end{spacing}

\noindent where $\mathbb{O}_{N_{\gamma},N_{\beta}}$ is the $N_{\gamma}\times N_{\beta}$
null matrix. According to Eq.~(6) of the main text, if $\mathfrak{p}\left(i\right)=\mathfrak{p}\left(j\right)=\gamma$
we get:

\begin{spacing}{0.8}
\begin{center}
{\small{}
\begin{align}
\mathrm{Cov}\left(V_{i}\left(t\right),V_{j}\left(t\right)\right)\approx & \frac{1-e^{-2\left|\lambda_{\gamma}^{\mathcal{P}}\right|t}}{2\left|\lambda_{\gamma}^{\mathcal{P}}\right|}\left(\sigma_{\gamma}^{\mathscr{B}}\right)^{2}\left(-\frac{1}{N_{\gamma}}\right),\;\mathrm{for}\;i\neq j\label{eq:branching-point-covariance}\\
\nonumber \\
\mathrm{Var}\left(V_{i}\left(t\right)\right)=\mathrm{Var}\left(V_{j}\left(t\right)\right)\approx & \frac{1-e^{-2\left|\lambda_{\gamma}^{\mathcal{P}}\right|t}}{2\left|\lambda_{\gamma}^{\mathcal{P}}\right|}\left(\sigma_{\gamma}^{\mathscr{B}}\right)^{2}\left(1-\frac{1}{N_{\gamma}}\right)\label{eq:branching-point-variance}
\end{align}
}
\par\end{center}{\small \par}
\end{spacing}

\noindent thus for $t\gg\frac{1}{\left|\lambda_{\gamma}^{\mathcal{P}}\right|}$
the amplitude of the fluctuations diverges for $\lambda_{\gamma}^{\mathcal{P}}\rightarrow0^{-}$,
while $\mathrm{Corr}\left(V_{i}\left(t\right),V_{j}\left(t\right)\right)\approx\frac{1}{1-N_{\gamma}}$.
Interestingly, according to \cite{Fasoli2015} this is the lower bound
of the correlation between fully-connected neurons in a homogeneous
population with size $N_{\gamma}$. Since $\frac{1}{1-N_{\gamma}}<0$
for $N_{\gamma}\geq2$, the neurons in population $\gamma$ are maximally
anti-correlated at the branching-point bifurcations. Moreover, according
to this formula, correlation tends to $-1$ only for $N_{\gamma}=2$
(which is confirmed by the top-right panel of Fig.~(4) in the main
text, in the special case $\mathfrak{P}=2$).

To conclude, if $\mathfrak{p}\left(i\right)=\mathfrak{p}\left(j\right)\neq\gamma$,
the matrix $\varPhi_{\mathfrak{p}\left(i\right),\mathfrak{p}\left(i\right)}\left(t\right)$
is not dominated by the term $e^{-\left|\lambda_{\gamma}^{\mathcal{P}}\right|t}$,
therefore we do not observe neither the divergence of the variance
nor close-to-one correlations in population $\mathfrak{p}\left(i\right)$.
This prevents also the formation of strong inter-population correlations
between $\gamma$ and $\mathfrak{p}\left(i\right)$. In the case $\mathfrak{P}=2$
this phenomenon occurs for the excitatory neurons since $\gamma=I$
and $\mathfrak{p}\left(i\right)=E$, and indeed this result is confirmed
by the top panels of Fig.~(4) in the main text.

\bibliographystyle{plain}
\bibliography{Bibliography}